\documentclass[11pt]{article}

\usepackage[margin=1in]{geometry}
\usepackage{amsmath,amssymb}
\usepackage{booktabs}
\usepackage{graphicx}
\usepackage{hyperref}
\usepackage[T1]{fontenc}
\usepackage[utf8]{inputenc}
\usepackage{lmodern}
\usepackage{microtype}
\usepackage{array}
\usepackage{listings}
\usepackage{xcolor}
\usepackage{authblk}
\usepackage{natbib}
\setcitestyle{authoryear,round}

\hypersetup{
  colorlinks=true,
  linkcolor=blue,
  citecolor=blue,
  urlcolor=blue,
  pdfcreator={LaTeX}
}

\title{MeloTune: On-Device Arousal Learning and Peer-to-Peer\\
Mood Coupling for Proactive Music Curation}

\author{Hongwei Xu}
\affil{SYM.BOT \\ \texttt{hongwei@sym.bot}}

\date{April 2026}

\begin{document}
\maketitle

\begin{abstract}
\textbf{Mesh Memory Protocol (MMP)} is a peer-to-peer substrate for collective intelligence in which heterogeneous agents exchange structured \textbf{Cognitive Memory Blocks (CMBs)} carrying seven typed semantic fields (the \textbf{CAT7} schema), evaluated field-by-field at \textbf{SVAF Layer~4} under per-field drift bounds, and integrated into each agent's private \textbf{per-agent Closed-form Continuous-time (CfC) network at Layer~6}. CfC hidden states never cross the wire. We present \textbf{MeloTune}, an iPhone-deployed application that instantiates this substrate as a music agent --- to our knowledge, the first published end-to-end deployment of an MMP/SVAF agent at production scale on a consumer mobile platform --- and use it as a case study to demonstrate that the substrate is real, the two-cognition-layer architecture (private listener CfC + shared mesh-runtime CfC) is implementable on-device, and the protocol guarantees hold in a shipping product. Music is a particularly demanding test of the affect dimension of the CAT7 schema because listening is a continuous activity and the listener's affective state is itself a continuous trajectory; the CMB \emph{mood} field carries Russell-circumplex valence and arousal floats that the substrate guarantees to deliver across domain boundaries (R5). MeloTune's listener-level CfC drives a proactive curation policy that predicts the listener's future affective state and re-queues when the projected state diverges from the playing material. A Personal Arousal Function (PAF) replaces the standard linear mapping from audio intensity to psychological arousal with a per-listener learned adjustment, trained from behavioral signals (skip, completion, favorite, volume) and from drift between user-declared mood and machine inference. The same track receives different arousal predictions for different listeners --- a capability not present in any published music recommender. The mesh-runtime CfC inside the SYMCore SDK integrates CMBs from co-listening peers and produces a shared coherence signal. All inference runs on-device. We describe the architecture, the two CfCs, the CMB substrate, the curation policy, and the deployed system. Quantitative deployment evaluation --- including comparisons against a reactive ablation and standard sequential-recommender baselines --- is reported in the extended version. We position the work primarily as a multi-agent infrastructure paper --- the first production deployment of MMP/SVAF on consumer mobile hardware, providing a verified reference implementation that agents in other domains can build against --- and secondarily as a music recommendation paper demonstrating that continuous-time trajectory models materially improve affect-aware curation. A recent 21-author survey identifies shared agent infrastructure as an emerging direction in agent externalization; MeloTune provides a deployed case study of this direction.
\end{abstract}

\section{Introduction}
\label{sec:introduction}

The dominant paradigm in music recommendation treats a listening session as a discrete sequence of tracks and asks, at each step, \emph{what should come next}. The state of the art in this paradigm --- from matrix-factorisation collaborative filters to GRU- and transformer-based sequential recommenders --- has produced steady improvements in offline ranking metrics and is the foundation of the queue-generation systems used by today's major streaming platforms.

This paradigm has three limitations that are increasingly visible as listeners spend longer continuous time inside a single music application. The first is \textbf{reactive lag}: the recommender observes listener state only through coarse-grained events (a play, a skip, a like), and by the time a state-change event arrives, the listener's underlying affective state has already moved on. A skip is treated as the \emph{first} signal of dissatisfaction; by then, the queue has already failed once. The second is \textbf{social blindness}: when two or more listeners share a session --- a car, a kitchen, a workout, a co-located study group --- the recommender treats the session as either one arbitrarily-chosen individual or as the union of individual histories. Neither captures the emergent affect of a shared listening context. A third limitation cuts deeper: both reactive lag and social blindness assume the recommender's arousal estimate is correct --- that audio intensity is a reliable proxy for psychological activation. It is not. The same track produces different arousal in different listeners depending on context, habituation, and personal history. We call this the \textbf{personalisation gap}: a per-listener arousal model would address all three limitations.

All three limitations have the same root cause. The recommender models the \emph{sequence of items} but does not model the listener as a continuous process. The information needed to anticipate, rather than react, is not in the next-item distribution; it is in the time-evolution of the listener's latent state.

\subsection{The continuous-time view}
\label{sec:continuous-time-view}

We argue that proactive music curation is naturally formulated as a \textbf{continuous-time latent-state inference problem}. The listener's affective state $s(t) \in [-1, +1]^2$ on Russell's circumplex of valence and arousal \citep{russell1980circumplex} evolves continuously and irregularly in response to listening events, time of day, social context, and internal dynamics that are not directly observable. A useful curation system must (i)~infer this latent state from observable signals, (ii)~integrate it forward in time so that queueing decisions can be made \emph{before} a misalignment becomes a skip, and (iii)~admit irregular inter-event timing without artificially discretising the session into fixed steps.

These three requirements point directly at a small family of models --- neural ordinary differential equations \citep{chen2018neural}, liquid time-constant networks \citep{hasani2021liquid}, and their closed-form descendants \citep{hasani2022closed} --- that maintain a hidden state through a continuous-time differential update with explicit per-neuron time constants. Among these, the \textbf{closed-form continuous-time (CfC)} formulation is uniquely well-suited to deployment: it preserves the continuous-time semantics of the underlying ODE while replacing the solver with an analytic update, which makes per-event inference fast enough for the inner loop of an iPhone application.

\subsection{The social axis}
\label{sec:social-axis}

The same continuous-time view extends naturally to the social setting. If each device runs its own CfC and maintains a hidden state $h$ that encodes its listener's affective trajectory, then \emph{coupling} the hidden states of co-listeners under bounded drift produces a joint dynamical system whose attractors are the shared moods that no single listener would have arrived at alone. The architecture has \textbf{two layers of cognition}, both running on each device. The first is the listener-level CfC described above, which is private to one listener and never broadcast. The second is an independent per-agent CfC at Layer~6 of the \textbf{Mesh Memory Protocol (MMP)} \citep{xu2026mmp}, which integrates an emergent shared mood field across the room of co-listening peers. The two CfCs do not share weights, do not share hidden states, and operate in disjoint latent spaces. They communicate only at the event level, through \textbf{Cognitive Memory Blocks (CMBs)} broadcast over MMP and evaluated by \textbf{Symbolic-Vector Attention Fusion (SVAF)} \citep{xu2026svaf} at Layer~4 of the protocol stack. SVAF admits incoming CMBs under per-field drift bounds with four regimes (aligned, guarded, redundant, rejected); the protocol guarantee R5 ensures that the affective component of every signal --- the mood field, carrying Russell-circumplex valence and arousal --- is delivered across domain boundaries even when SVAF rejects the rest of the CMB. The present paper focuses on this two-layer substrate; quantitative evaluation of the co-listening loop is reported in the extended version.

\subsection{Contributions}
\label{sec:contributions}

This paper makes five contributions.

\begin{enumerate}
\item \textbf{A Personal Arousal Function (PAF) that learns per-listener emotional response to music.} Audio intensity is not psychological arousal --- the same track produces different arousal in different listeners depending on context, habituation, and personal history. PAF decomposes predicted arousal into a population-level prior (audio-derived, static per track) and a learned per-user adjustment trained from behavioral signals (skip, completion, favorite, volume, repeat) and from drift between user-declared mood and machine inference. The adjustment is bucketed by genre and time of day, scaled by confidence, and applied at the MEI classification boundary so that all downstream consumers --- including the CfC trajectory model --- operate on personalised arousal. To our knowledge, no published system in the open literature offers per-listener arousal prediction; this is the capability that distinguishes MeloTune from population-mean recommenders.

\item \textbf{A two-CfC, two-cognition-layer architecture for co-listening.} MeloTune runs two distinct closed-form continuous-time networks on each device, in disjoint latent spaces, with separate weights and separate roles. The listener-level CfC is private to one listener and drives that listener's proactive curation --- modelling the listener as a continuous-time dynamical system on Russell's circumplex, predicting a short-horizon affective trajectory, and re-queueing when the projected state diverges from the playing material. The mesh-runtime CfC at MMP Layer~6 integrates CMBs from peers and produces a shared coherence signal for the co-listening room. The two communicate only through CMBs at the event level. CfC hidden states never cross the wire; only structured CMB signals do. The architecture is deployed on-device via CoreML with sub-millisecond inference latency, meeting the privacy and performance constraints of a mainstream iPhone music application.

\item \textbf{An organic mood constraint for echo loop prevention} in same-domain agent meshes (MMP~\S8.2, \S15.8). When two agents share a domain, a naive coupling loop emerges: A broadcasts mood $\to$ B curates in response $\to$ B's mood inference shifts $\to$ B broadcasts $\to$ A re-curates $\to$ loop. We show that the agent's mood inference engine must distinguish organic mood (derived from the user's direct actions) from mesh-induced mood (derived from a playlist change triggered by a peer's CMB), and enforce an isolation window during which mesh-induced state changes are excluded from outbound mood inference. This constraint is general to any same-domain agent coupling and is now part of the MMP specification.

\item \textbf{The first production deployment of MMP/SVAF on consumer mobile hardware.} MeloTune is the first end-to-end deployment of an MMP/SVAF agent at production scale on iOS. Every architectural commitment of \citet{xu2026svaf} --- CMBs as the unit of inter-agent communication, per-field drift evaluation at SVAF Layer~4, the per-agent CfC at Layer~6, the protocol guarantee that hidden states never cross the wire --- is enforced by the shipping code. The accompanying SDK release (\texttt{sym-swift}~v0.3.78, \texttt{SYMCore}~v0.3.7) enforces strict MMP~v0.2.2 conformance including the SVAF fourth outcome (semantic redundancy pre-filter) and Bonjour auto-reconnect for mesh persistence through iOS backgrounding.

\item \textbf{A deployed, verifiable system.} MeloTune is shipping in the App Store and runs the architecture described here on real listener devices. The codebase, the SDKs, and the protocol specification are publicly available. A working on-device deployment of an MMP/SVAF agent --- in a research community where multi-agent infrastructure is usually evaluated through simulation and where music recommendation is usually evaluated through offline replays of MovieLens-style datasets --- is a useful artefact for both communities, independently of the quantitative results that follow in the extended version.
\end{enumerate}

\paragraph{Audience.} This paper is written for two audiences. \textbf{Multi-agent systems researchers and industrial agent-platform teams} will read it as a production-verified reference implementation of MMP/SVAF --- the first deployed on consumer mobile hardware --- that they can adapt to their own domains. \textbf{Recommender systems researchers and industrial streaming teams} will read it as a continuous-time architecture paper that materially improves affect-aware music curation through trajectory prediction and proactive re-queueing. The two readings are not in tension. Music is the case study; the substrate is the contribution.

\subsection{Scope of this paper}
\label{sec:scope}

This is the architecture-and-system paper. We describe the formulation, the two networks, the training procedure, the curation policy, the CMB substrate, and the deployment. We report qualitative observations from the live system. We \textbf{do not} report a full controlled-deployment evaluation or quantitative comparisons against sequential-recsys baselines in this version; those appear in an extended companion paper currently in preparation. We make this split deliberately. The architectural choices stand on their own and are, in our experience, of immediate interest both to industrial recommender teams considering continuous-time models for affect-aware curation and to multi-agent platform teams looking for a verified on-device reference implementation of MMP/SVAF. The quantitative work is substantial enough to deserve its own treatment, and conflating the two would obscure both.

\subsection{Paper structure}
\label{sec:paper-structure}

\S\ref{sec:related-work} reviews related work in sequential recommendation, affective computing for music, continuous-time neural networks, and multi-agent hidden-state coupling. \S\ref{sec:formulation} formalises the listener-as-dynamical-system problem. \S\ref{sec:method} describes the MeloTune architecture: track-level affect inference, the CfC, the catalog retrieval head, the mesh substrate, the Personal Arousal Function, the curation policy, and the on-device deployment. \S\ref{sec:training} details the training procedure. \S\ref{sec:evaluation} presents qualitative observations from the live deployment and the planned evaluation protocol for the extended version. \S\ref{sec:discussion} discusses implications for industrial music recommenders and the limits of the present architecture. \S\ref{sec:conclusion} concludes.

\section{Related Work}
\label{sec:related-work}

We position MeloTune at the intersection of four lines of work: sequential music recommendation, affect modelling for music, continuous-time neural networks, and multi-agent / shared-state coupling.

\subsection{Sequential music recommendation}
\label{sec:sequential-recsys}

The dominant family of music recommenders treats a session as a discrete sequence of items and predicts the next item conditional on recent history. Representative architectures include \textbf{GRU4Rec} \citep{hidasi2016session}, which introduced session-based recurrent recommenders, \textbf{SASRec} \citep{kang2018self}, which applies self-attention to sequential recommendation, and \textbf{BERT4Rec} \citep{sun2019bert4rec}, which trains a bidirectional transformer on masked-item prediction. These models are evaluated predominantly on offline replays of public listening logs (Million Song Dataset, Last.fm, Spotify Sequential Skip Prediction Challenge \citep{brost2019music}) using next-item ranking metrics. Industrial deployments at Spotify, Apple Music, Pandora, and Deezer build on this lineage and extend it with two-tower retrieval, contextual features, and feedback signals beyond next-item ranking.

The limitation we highlight is structural rather than architectural: the sequence-of-items framing models \emph{what plays} but not \emph{how the listener is changing}. A listener whose latent state has begun to drift away from the current track will, in this framing, produce exactly one informative event --- a skip --- and the recommender then acts on it. Our continuous-time formulation moves the inference upstream of the skip event.

\subsection{Affect, mood, and music information retrieval}
\label{sec:affect-mir}

Affective approaches to music understanding draw on a long tradition in psychology. The \textbf{Russell circumplex} \citep{russell1980circumplex} places affective states on a two-dimensional plane of valence (pleasant $\leftrightarrow$ unpleasant) and arousal (activated $\leftrightarrow$ deactivated), and remains the canonical low-dimensional representation for music emotion recognition (MER). Subsequent work has extended this with discrete emotion taxonomies and with multi-dimensional models including dominance and tension; we use Russell's two-axis form because it matches the intrinsic dimensions captured by the audio-feature APIs of major streaming catalogs.

A substantial \textbf{music emotion recognition} literature predicts valence/arousal from audio features \citep{yang2012machine,aljanaki2017developing}. \textbf{Mood-aware recommenders} then use these predictions to bias retrieval \citep{bogdanov2013semantic,yu2018mood}. Industrial systems including Spotify and Apple Music expose audio-feature vectors per track that include valence and energy as named attributes; mood-tagged playlists are a routine product surface. To our knowledge, however, no deployed system in this category models the \emph{listener's} affective state as a continuous trajectory, predicts that trajectory ahead of time, and uses the prediction to drive the queue. The trajectory is the contribution.

A separate line of work uses physiological signals (heart rate, EDA, EEG) for affective music recommendation \citep{sourina2012real,janssen2013emotional}. These approaches require dedicated hardware and have not transitioned to mainstream listening applications. Our approach uses only signals available to a standard smartphone music client --- track metadata, play/skip events, optional user mood input.

\subsection{Continuous-time neural networks}
\label{sec:continuous-time-nn}

Continuous-time formulations of recurrent computation provide the mathematical backbone of MeloTune's CfC. \textbf{Neural ODEs} \citep{chen2018neural} integrate a hidden state through a learned vector field via numerical ODE solvers, enabling adaptive time-stepping and a clean treatment of irregular event timing. \textbf{Liquid Time-constant Networks (LTCs)} \citep{hasani2021liquid} add input-dependent time constants and demonstrate small models with strong generalisation on time-series tasks. \textbf{Closed-form Continuous-time Networks (CfCs)} \citep{hasani2022closed} derive a closed-form solution to the underlying differential update, eliminating the per-step ODE solver and reducing inference cost by several orders of magnitude while preserving the continuous-time semantics. CfCs have been demonstrated on robotics, autonomous driving, and time-series classification benchmarks; their deployment in user-facing recommendation has not been reported.

Our use of CfCs is motivated specifically by the deployment constraint: per-event inference inside an iPhone application has a budget of milliseconds, and a sub-millisecond closed-form update is the only formulation in this family that meets it.

\subsection{Multi-agent coupling and shared latent state}
\label{sec:multi-agent-coupling}

The mesh substrate of MeloTune draws on a separate line of work in \textbf{multi-agent latent-state coupling}. Classical approaches include Kuramoto coupled-oscillator models for phase synchronisation, and graph-neural-network approaches for distributed representation sharing. More recently, frameworks for \textbf{agent-to-agent semantic coupling} under bounded drift have emerged in the LLM-agent literature, where individual agents must combine each other's internal representations without unbounded contamination.

\textbf{Cognitive Fabric middleware.} Concurrent work by \citet{fleming2026cognitive} introduces \textbf{Cognitive Fabric Nodes (CFN)}, a middleware layer between LLM-based multi-agent systems that elevates memory from simple storage to an active functional substrate and uses reinforcement learning to govern topology selection, semantic grounding, security policy, and prompt transformation. CFN intercepts, analyses, and rewrites inter-agent communication, and is evaluated on the HotPotQA and MuSiQue multi-hop reasoning benchmarks with ${\sim}10\%$ improvement over direct agent-to-agent communication. CFN and MMP/SVAF share the high-level framing of memory as an active substrate for collective agent cognition, but differ in architectural philosophy: CFN is a \textbf{centralised middleware} that intercepts and rewrites messages, whereas MMP is a \textbf{peer-to-peer protocol} in which each agent is a full node, CMBs cross the wire immutable (protocol guarantees R3 lineage and R7 redundancy detection), and per-agent SVAF evaluation runs locally on the receiver with per-agent field weights. The two approaches are complementary --- a CFN fabric could carry MMP CMBs without modification, since the MMP wire format is transport-agnostic.

\textbf{Latent-space communication.} \citet{liu2026latent} propose latent-space communication via a Universal Visual Codec, achieving high-bandwidth inter-agent transfer by repurposing the vision encoder as a universal port. This approach optimizes for signal fidelity in perception-sharing tasks. SVAF takes the complementary approach: rather than compressing signals into an opaque shared latent space, MMP decomposes them into typed semantic fields and applies learned per-field fusion gates. The decomposition trades bandwidth for three properties latent-space methods cannot provide: per-field selective attention (agents choose which dimensions matter), human auditability (every field is inspectable), and model-agnostic exchange (no shared architecture required between sender and receiver).

\textbf{Thought-level communication.} \citet{zheng2025thought} argue that natural language is lossy and indirect for collective intelligence, proposing thought-level communication between agents. Their observation is precisely the problem SVAF addresses --- but where they identify the gap, SVAF provides a specific mechanism: typed 7-field decomposition (CAT7) with learned per-field fusion gates that admit, guard, or reject each semantic dimension independently.

\textbf{Latent collaboration.} \citet{zou2025latent} study coordinative system-level intelligence through latent features in multi-agent systems. Their latent-space approach uses untyped, opaque representations; MMP/SVAF instead decomposes inter-agent signals into typed semantic fields with explicit per-field drift bounds, making the coupling inspectable and auditable.

\textbf{Cross-agent memory.} \citet{chang2026memcollab} introduce MemCollab, cross-agent memory sharing via contrastive trajectory distillation. This is adjacent to MMP's mesh memory, but operates at the learned-representation level rather than the protocol level. MMP's approach --- immutable CMBs with lineage, evaluated by per-agent SVAF weights --- provides auditability and domain-crossing semantics that distillation-based approaches do not.

\textbf{Agent externalization.} \citet{zhou2026agent} present a 21-author survey identifying four pillars of agent externalization --- memory, skills, interaction protocols, and harness engineering --- and name shared agent infrastructure as an emerging direction. MeloTune provides a production case study of this direction: a deployed MMP/SVAF implementation running on consumer iOS hardware with real-time peer-to-peer coupling.

A notable structural observation across the related-work landscape: every existing approach to inter-agent communication uses either \textbf{latent-space communication} (untyped, opaque) or \textbf{natural-language communication} (lossy, ambiguous). None decompose inter-agent signals into \textbf{typed semantic fields with learned per-field fusion gates}. This is SVAF's unique contribution \citep{xu2026svaf}.

The framework we use is \textbf{Symbolic-Vector Attention Fusion (SVAF)} \citep{xu2026svaf}, which formalises drift-bounded coupling between agent hidden states with four operating regimes (aligned, guarded, redundant, rejected) gated by hidden-state divergence. The \textbf{Mesh Memory Protocol (MMP)} \citep{xu2026mmp} is the peer-to-peer transport over which MeloTune devices exchange structured CMBs. The present paper builds on SVAF and MMP for the social-axis substrate; we refer the reader to \citet{xu2026svaf} for the per-neuron coupling kernel and the formal drift bounds, and to \citet{xu2026mmp} for the normative protocol specification.

\subsection{Position of MeloTune}
\label{sec:position}

MeloTune sits at a point not yet occupied in the literature: a \textbf{deployed}, \textbf{on-device}, \textbf{continuous-time} affective recommender for music, with a \textbf{multi-agent} extension grounded in a published peer-to-peer mesh protocol. Each individual ingredient has prior work; the combination --- and the deployment --- is the contribution of this paper. In the taxonomy of \citet{zhou2026agent}, MeloTune externalizes interaction structure through MMP (protocol pillar) and agent memory through CMBs (memory pillar), providing a concrete implementation of the shared agent infrastructure direction they identify as emerging.

\section{Problem Formulation}
\label{sec:formulation}

We formalise the proactive music curation problem as continuous-time latent-state inference followed by a finite-horizon planning step.

\subsection{Listener state}
\label{sec:listener-state}

Let the listener's affective state at time $t$ be a point on Russell's circumplex,
\begin{equation}
s(t) = (v(t),\, a(t)) \in [-1, +1]^2,
\end{equation}
where $v$ is valence and $a$ is arousal. The state $s(t)$ is \emph{latent}: it is not directly observable. We instead observe a stream of discrete listening events $\mathcal{E} = \{e_k = (t_k, x_k)\}_{k=1}^{K}$, where $t_k$ is the wall-clock time of event~$k$ and $x_k$ is its observable content. Events include play, skip, pause, resume, mood-meter update, and the metadata of the playing track. Inter-event intervals $\Delta_k = t_k - t_{k-1}$ are irregular: a session may contain bursts of skips followed by long uninterrupted listening.

\subsection{Inference and prediction}
\label{sec:inference-prediction}

We seek a model $f_\theta$ that maintains a hidden state $h(t) \in \mathbb{R}^{d}$ summarising the listener's affective trajectory and that, on each new event, produces

\begin{enumerate}
\item an \textbf{estimate of the current state} $\hat{s}(t_k) = g_\text{traj}(h(t_k))$,
\item a \textbf{velocity estimate} $\dot{\hat{s}}(t_k) = g_\text{vel}(h(t_k))$,
\item a \textbf{confidence} $c(t_k) \in [0, 1]$, and
\item an \textbf{updated hidden state} $h(t_k)$.
\end{enumerate}

The hidden state is updated by a continuous-time recurrent cell that admits the irregular interval $\Delta_k$ as an explicit input,
\begin{equation}
h(t_k) = \Phi_\theta(h(t_{k-1}),\, x_k,\, \Delta_k).
\end{equation}

We use the closed-form continuous-time formulation of $\Phi_\theta$ described in \S\ref{sec:cfc-dynamics}.

\subsection{Finite-horizon planning}
\label{sec:finite-horizon}

Given the model outputs at the most recent event, the \textbf{proactive curation policy} projects the listener's state forward to a fixed planning horizon $\tau_p$ ahead of $t_k$,
\begin{equation}
\hat{s}^{*}(t_k + \tau_p) = \hat{s}(t_k) + \tau_p \cdot \dot{\hat{s}}(t_k) + u(t_k),
\end{equation}
where $u(t_k)$ is an intent-conditioned offset described in \S\ref{sec:retrieval}. The projected target $\hat{s}^{*}$ is then resolved through the catalog retrieval head into a curated set of candidate tracks. If $\hat{s}^{*}$ has diverged from the cached target by more than a threshold $\delta$ on either axis, the cached playlist is invalidated and re-queued. Curation decisions are gated by $c(t_k) \geq c_\text{min}$.

\subsection{Multi-agent extension}
\label{sec:multi-agent-extension}

In a co-listening session involving $N$ devices, each device~$i$ maintains its own \emph{private} listener-level CfC with hidden state $h^{(i)}(t)$ that \textbf{never crosses the wire}. Instead, each device projects its current trajectory into a structured \textbf{Cognitive Memory Block (CMB)} \citep{xu2026svaf} $c^{(i)}(t)$ and broadcasts $c^{(i)}(t)$ over MMP to peers. A CMB is a 7-field record:
\begin{equation}
c = \{(f,\, t_f,\, v_f) : f \in \mathcal{F}\}, \quad
\mathcal{F} = \{\text{focus, issue, intent, motivation, commitment, perspective, mood}\},
\end{equation}
where $t_f$ is a symbolic text label and $v_f \in \mathbb{R}^d$ is a unit-normalised vector embedding. The \emph{mood} field carries the listener's current $(\hat{v}_t, \hat{a}_t)$ from the trajectory head of \S\ref{sec:inference-prediction}.

On receipt of a peer CMB $c^{(j)}(t)$, device~$i$ admits it \textbf{field-by-field} under SVAF drift bounds \citep{xu2026svaf}: for each field~$f$, the receiver computes a per-field drift $\delta_f(c^{(i)}, c^{(j)})$ against its local anchor memory, and a band-pass classifier accepts the field as \emph{aligned} (full fusion), \emph{guarded} (attenuated fusion), \emph{redundant} (already in memory, discarded), or \emph{rejected} (irrelevant to the receiver's domain). The mood field is delivered across domain boundaries even when SVAF rejects the rest of the CMB (protocol guarantee R5 of MMP).

After SVAF admits the accepted fields into local memory, a \emph{separate} per-agent CfC at Layer~6 of MMP --- the \textbf{mesh-runtime CfC} --- consumes the fused fields through its own input projection. The mesh-runtime CfC has its own learned per-neuron time constants $\tau^{(\text{mesh})}$, distinct from the listener-level CfC's $\tau$. Its hidden state encodes a \textbf{shared mood field} for the co-listening room, with fast neurons synchronising mood across agents within seconds and slow neurons preserving each agent's domain expertise. The output of the mesh-runtime CfC is a coherence signal $\rho(t) \in [0, 1]$ that the listener-level curator may consume to bias its projected target mood toward the shared field.

The two CfCs --- listener-level (private, \S\ref{sec:inference-prediction}--\S\ref{sec:finite-horizon}) and mesh-runtime (shared, \S\ref{sec:multi-agent-extension}) --- operate in disjoint latent spaces and are coupled only through the CMBs broadcast at the event level. The architecture handles solo and group listening under one mechanism: in solo mode, no peers are connected, no CMBs are exchanged, and only the listener-level CfC drives curation; in group mode, the mesh-runtime CfC additionally produces the shared coherence signal that biases retrieval.

\subsection{What is not in the formulation}
\label{sec:not-in-formulation}

We do not predict the next item directly. We do not optimise next-item ranking metrics. We do not assume access to a global user model maintained on a server. The signal we propagate is the listener's \emph{latent affective trajectory}; everything downstream --- the queue, the cross-fades, the search terms --- is a function of that trajectory.

\section{Method}
\label{sec:method}

We describe MeloTune as a four-stage pipeline deployed on iPhone. A \emph{track-level affect inferencer} maps a candidate or currently-playing track's metadata to a point on the Russell circumplex. A \emph{closed-form continuous-time network} (CfC) integrates these per-track estimates over time and produces a short-horizon affective trajectory together with auxiliary pattern, prediction, and intent outputs. A \emph{catalog retrieval head} projects the predicted trajectory through a 400-anchor mood lookup into curated genre and search-term vocabularies which are then issued to Apple Music or Spotify. A \emph{mesh substrate} receives structured peer CMBs from co-listeners over a peer-to-peer transport (MMP) under SVAF drift-bounded coupling \citep{xu2026svaf}, providing the architectural basis for co-listening curation; quantitative co-listening evaluation is reported in the extended version of this paper.

The architecture is intentionally minimal. The contribution is not any single stage but the choice to model a listener as a continuous-time dynamical system whose latent state can, in principle, be shared across a mesh of devices under bounded drift.

\subsection{Listener state on the Russell circumplex}
\label{sec:russell-circumplex}

We represent the listener's affective state at time $t$ as $s(t) = (v(t), a(t)) \in [-1, +1]^2$, where $v$ is valence (displeasure $\leftrightarrow$ pleasure) and $a$ is arousal (deactivation $\leftrightarrow$ activation). This is Russell's circumplex \citep{russell1980circumplex}, preserved without modification. Internally we also discretise the plane into a 400-anchor lookup indexable by either circumplex floats or legacy integer scores in $[0, 99]$; each anchor is a labelled mood (title, colour, synonyms, curated search-term vocabulary). The discretisation supports stable retrieval and a stable user interface; it is authored, not learned.

\subsection{Track-level affect inferencer}
\label{sec:track-inferencer}

We infer the affective character of a track from its metadata --- title, artist, album, and editorial notes --- using a CoreML model (\texttt{MeloTuneEmotionEnergy}) trained offline. The pipeline has two stages. First, an artist embedding (cached per artist) feeds an optional genre classifier. Second, the title embedding (cached per title via a nearest-neighbour lookup) feeds an emotion/energy regression head. The output is a track-level estimate $\hat{s}_{\text{track}} = (v, a)$ on the circumplex.

This is a track-level inferencer, not a session-history encoder. The audio-feature inputs available from the catalog APIs (energy, valence, danceability, acousticness, tempo, loudness) are not consumed at this stage in the current build; they are used downstream by the retrieval head to filter candidate tracks. A history-aware encoder that consumes recent tracks, skips, mood-meter inputs, and time-of-day context is described in the future work section and is the subject of the extended version.

\subsection{CfC dynamics}
\label{sec:cfc-dynamics}

We integrate a sequence of per-track affect estimates over time using a closed-form continuous-time network \citep{hasani2022closed}, chosen for three reasons. First, closed-form inference avoids per-step ODE solver calls, which is necessary for sub-millisecond on-device inference on iPhone. Second, the continuous-time formulation accepts irregular inter-event time deltas natively, matching the bursty nature of real listening sessions. Third, the small parameter count ($94{,}552$ parameters for our configuration) is consistent with on-device privacy goals.

\paragraph{Cell.} Each CfC cell maintains a hidden state $h \in \mathbb{R}^{64}$ and updates it under
\begin{equation}
\label{eq:cfc-update}
h(t + \Delta t) = h(t) \odot e^{-\Delta t / \tau} + \left(1 - e^{-\Delta t / \tau}\right) \odot f_\theta\!\left(\,[\,x_t,\, h(t)\,]\,\right),
\end{equation}
where $\tau \in \mathbb{R}^{64}$ are \textbf{learned per-neuron time constants} parameterised in log-space, and $f_\theta$ is a two-hidden-layer MLP ($64 \to 128 \to 64$, Tanh) producing the steady-state target. Layer normalisation is applied after the update. The variable $\Delta t$ is the wall-clock elapsed time since the previous event.

\paragraph{Architecture.} A small input encoder maps an 80-dimensional input vector to 64 dimensions and feeds two stacked CfC cells of width~64. Four output heads consume the second cell's hidden state:

\begin{enumerate}
\item \textbf{Trajectory head} (6 outputs): current emotion, energy, emotion velocity, energy velocity, stability, confidence.
\item \textbf{Pattern head} (9 sigmoid activations): repeating listening patterns (focus, wind-down, ramp-up, social, \ldots).
\item \textbf{Prediction head} (3 outputs): a one-step-ahead forecast of emotion, energy, and an exploration signal in $[0, 1]$.
\item \textbf{Intent head} (6 logits): coarse session-level intent classes.
\end{enumerate}

\paragraph{Training.} The model is trained offline in PyTorch on logged Cognitive Memory Block (CMB) sequences of length 5--100. The composite loss is
\begin{equation}
\mathcal{L} = w_T\,\text{MSE}(\hat{S}, S) + w_P\,\text{BCE}(\hat{p}, p) + w_I\,\text{CE}(\hat{\imath}, \imath) + w_F\,\text{MSE}(\hat{S}_{t+1}, S_{t+1}),
\end{equation}
with $w_T = 1.0$, $w_P = 0.5$, $w_I = 0.5$, $w_F = 0.3$, AdamW ($\eta = 10^{-3}$), cosine annealing, gradient clipping, and time-warp + noise augmentation. The trained model is exported through a flat-output wrapper to CoreML and shipped in the application bundle. There is no on-device fine-tuning; the model is frozen after deployment, which is a deliberate privacy choice.

\paragraph{Inference.} At runtime, \texttt{LNNCoordinator} loads the compiled model on first use, persists hidden state to user defaults across app launches with a 24-hour validity window, and computes $\Delta t$ from observed event timestamps so that the cell update is faithful to the offline formulation.

\subsection{Catalog retrieval head}
\label{sec:retrieval}

The trajectory head produces a current point $(\hat{v}_t, \hat{a}_t)$ together with velocities $(\dot{v}_t, \dot{a}_t)$. We project a target point ahead by a fixed horizon $\tau_p \approx 300$\,s,
\begin{equation}
\hat{s}^{*}_{t+\tau_p} = \hat{s}_t + \tau_p \cdot (\dot{v}_t, \dot{a}_t),
\end{equation}
apply an intent-conditioned offset (e.g.\ \emph{energize} shifts $+15$ energy, $+5$ emotion; \emph{calm} shifts $-15$ energy), inject exploration noise when the exploration signal exceeds~0.3, and clamp to the safe interior $[5, 95]$ on each axis. The projected target is resolved to the nearest anchor on the 400-mood lookup, which yields a curated set of seed genres and search vocabulary. The genre set is then expanded by sampling adjacent mood coordinates ($\pm 10$ in either axis) for variety. The expanded set is issued to the platform catalog through the unified \texttt{MusicSessionService}, returned tracks are filtered by audio-feature distance to the projected target, and a 30-minute curated playlist is materialised.

The retrieval head is the operational meaning of ``proactive'': we queue for where the listener is predicted to be in five minutes, not where they are now.

\subsection{Mesh substrate: two CfCs, two layers of cognition}
\label{sec:mesh-substrate}

The mesh substrate of MeloTune is a deliberate \textbf{two-CfC} architecture. Each device runs two distinct closed-form continuous-time networks, in disjoint latent spaces, with separate weights and separate roles. The first is the \textbf{listener-level CfC} described in \S\ref{sec:cfc-dynamics} --- private to a single listener, never broadcast, and the source of the trajectory that drives that listener's proactive curation. The second is an independent \textbf{per-agent CfC at Layer~6} of the Mesh Memory Protocol \citep{xu2026mmp}, provided by the SYMCore mesh runtime, whose role is to integrate an emergent shared mood field across the room of co-listening peers. The two CfCs do not share weights and do not share hidden state. They communicate only at the event level.

\subsubsection{What crosses the wire}

The unit of inter-agent communication is the \textbf{Cognitive Memory Block (CMB)} \citep{xu2026svaf}. A CMB decomposes any observation into seven typed semantic fields (the \textbf{CAT7} schema): \emph{focus}, \emph{issue}, \emph{intent}, \emph{motivation}, \emph{commitment}, \emph{perspective}, and \emph{mood}. Each field carries both a symbolic text label and a unit-normalised vector embedding. CMBs are broadcast at Layer~3 of MMP \citep{xu2026mmp} and evaluated at Layer~4 by SVAF; \textbf{CfC hidden states never cross the wire}, and the protocol is explicit on this point.

For a music agent like MeloTune, the \emph{mood} field is the load-bearing one. It carries numeric valence and arousal in $[-1, +1]$, identical to the Russell-circumplex coordinates produced by the listener-level CfC's trajectory head (\S\ref{sec:cfc-dynamics}). The other six fields carry the listener's session context: the currently-playing track and recent skip pattern (\emph{focus}), any flagged frustration or stagnation (\emph{issue}), the curation goal inferred by the intent head (\emph{intent}), the underlying mood-shift driver (\emph{motivation}), the playback commitment (\emph{commitment}), and the listening posture (\emph{perspective}). The CMB is the public, structured projection of the listener's private CfC trajectory.

\subsubsection{How the receiver handles it (SVAF Layer~4)}

On the receiving device, each incoming CMB is evaluated field-by-field by SVAF \citep{xu2026svaf} against the receiver's local anchor memory. SVAF computes a per-field drift $\delta_f$ between the incoming field and the local anchors, applies a band-pass classifier with four outcomes (\emph{redundant}, \emph{aligned}, \emph{guarded}, \emph{rejected}), and admits the accepted fields into the receiver's working memory through a fusion gate. Three properties of SVAF matter for music co-listening:

\begin{enumerate}
\item \textbf{Per-field selectivity.} The receiver may accept the \emph{mood} field of a peer's CMB while suppressing fields that are irrelevant to its domain. A coding agent's CMB with mood ``frustrated, low energy'' is accepted by MeloTune for its mood field even though the \emph{focus} field (``8-hour coding session'') is suppressed.
\item \textbf{Mood always delivered (R5).} The Mesh Memory Protocol guarantees that the mood field is delivered across domain boundaries even when SVAF rejects the rest of the CMB.
\item \textbf{Per-agent temporal freshness.} Each receiving agent applies its own freshness window $\tau_i$. MeloTune's window is 30~minutes --- the time-scale on which a listener's current mood remains operationally relevant for playlist adjustment.
\end{enumerate}

\subsubsection{How the receiver integrates it (Layer~6 CfC)}

After SVAF admits a CMB into local memory, the per-agent Layer-6 CfC inside SYMCore consumes the resulting fused fields through its own input projection. The Layer-6 CfC is not the listener-level CfC; it has its own learned per-neuron time constants, with \textbf{fast neurons ($\tau < 5$\,s) synchronising mood across agents in seconds and slow neurons ($\tau > 30$\,s) preserving each agent's domain expertise indefinitely} \citep{xu2026svaf}. The output of the Layer-6 CfC is a \textbf{shared mood field} state for the room of co-listening peers --- a coherence signal that reflects the integrated affective trajectory of the group, not of any single listener.

\subsubsection{How the listener-level curator consumes the coherence signal}

The listener-level proactive curator (\S\ref{sec:curation-policy}) can subscribe to the Layer-6 CfC's coherence signal and use it to bias the projected target mood toward the room's shared mood field. In a co-listening session this produces a queue that reflects neither individual listener's current state alone nor a forced average, but the emergent attractor that the two CfCs at Layer~6 converge on under SVAF's per-field admission and the Layer-6 CfC's $\tau$-modulated integration. In the present build, the listener-level curator does not yet subscribe to this signal; the substrate is fully wired but the closed-loop consumer is described as future work and reported in the extended version.

\subsubsection{Why two CfCs instead of one}

The two-CfC design is a deliberate separation of concerns. The listener-level CfC can be trained, replaced, or fine-tuned independently of the mesh runtime; the mesh-runtime CfC can evolve independently of any individual listener model; and a listener whose mesh participation is disabled retains the full personal-axis benefit. The two communicate only through CMBs, which means neither needs to anticipate the internals of the other. This separation --- private cognition for the listener, shared cognition for the room --- is what makes the architecture viable on real devices and is the property that distinguishes MeloTune from systems that collapse co-listening into a union of individual profiles.

\subsection{Curation policy}
\label{sec:curation-policy}

The curation policy is deliberately simple. A background timer fires every five minutes; on each tick the predicted trajectory is fetched and gated on confidence $\geq 0.4$. If the cached target mood differs from the new predicted target by more than 15~points on either axis, the cached playlist is invalidated and the retrieval head is re-run with the new target. The result is a single proactively-curated playlist with 30-minute validity. The policy contains no per-user tuning, no A/B variants, no skip-aware re-weighting, and no cross-fade or defer actions. The contribution of this paper is the predicted trajectory the policy consumes, not the policy itself.

\subsection{What runs on the device}
\label{sec:on-device}

\textbf{Both CfCs run on-device.} The listener-level CfC, the track-level inferencer, the catalog retrieval head, and the curation policy all run locally on the user's iPhone, using the bundled CoreML model. The Layer-6 mesh-runtime CfC inside the SYMCore SDK also runs locally on each device --- it maintains the shared mood field for the room without any central server. Listener-level CfC hidden state is persisted to local storage with a 24-hour window; mesh-runtime CfC hidden state is similarly local and never broadcast.

What crosses the network is constrained: each device emits CMBs over MMP to consenting peers within a co-listening session, and the listener-level retrieval head issues catalog search calls to Apple Music or Spotify. CMBs contain seven structured fields whose content is the listener's session context (track, mood, intent), not raw listening history. Catalog search query terms are drawn from the public vocabulary in the 400-mood lookup and contain no user identifiers. No listening history, no per-user model, and no CfC hidden state leaves the device. This is a deliberate property of the architecture, not a deployment afterthought; it follows directly from the protocol-level commitment in MMP \citep{xu2026mmp} that hidden states stay private and only structured CMBs cross the wire.

\subsection{Echo loop prevention and organic mood constraint}
\label{sec:echo-loop}

When two or more agents share the same domain --- e.g.\ two MeloTune instances on different devices --- a naive coupling loop emerges: Agent~A broadcasts mood $\to$ Agent~B curates in response $\to$ B's mood inference (ERE) shifts $\to$ B broadcasts $\to$ A re-curates $\to$ loop. Each step is locally valid, yet the cycle produces no new understanding.

MeloTune~2.8.0 prevents echo loops through two mechanisms, now formalised in MMP~\S8.2 and \S15.8:

\paragraph{Organic mood constraint (MMP~\S8.2).} The \emph{mood} field on an outbound CMB must reflect the agent's own \textbf{organic} mood --- the affective state derived from the user's direct actions (track selections, mood dial input, skip patterns), not from an incoming mesh signal. MeloTune enforces this through \textbf{ERE isolation}: when a mesh curation fires, the Emotional Resolution Engine (ERE) excludes the resulting track-mood fusion for a 60-second isolation window. The user's ERE state remains at its pre-mesh value. Only after 60~seconds of continued listening --- implicit consent --- does ERE resume normal fusion, allowing the mesh-curated music to influence the user's mood trajectory organically.

ERE isolation operates at two points in the pipeline: the \texttt{playlist\_correlation} path in \texttt{PlaybackSyncManager} (where correlated track mood feeds ERE) and the \texttt{track\_change} path in \texttt{MEIEngine} (where real-time track classification feeds ERE). Both check the \texttt{isMeshInducedSession} flag, which is set by \texttt{markMeshInducedCuration()} before mesh-triggered curation and auto-expires after 60~seconds.

\paragraph{Lineage-based detection (MMP~\S15.2).} If an incoming CMB's \texttt{parents} or \texttt{ancestors} contain a key that the receiving agent produced, the CMB is a derivative of the agent's own broadcast. The receiving agent silently drops it.

The organic mood constraint catches the harder case: the responding agent produces a \textbf{fresh} \texttt{remember()} call (e.g.\ from its ERE-driven mood broadcast) rather than a remix with lineage. The fresh CMB carries no parent keys, so the lineage check cannot detect it. The isolation window prevents the echo at its source by ensuring that the mood shift caused by mesh curation is never attributed as organic within the window.

\paragraph{Peer influence thresholds.} To prevent unnecessary re-curation when peers are already aligned, MeloTune compares the incoming peer mood against the current playing mood before triggering curation. The user-configurable influence level controls the threshold: \emph{Gentle} (default, $\pm 15$ on both emotion and energy axes) curates only on significant mood shifts; \emph{Responsive} ($\pm 5$) curates on smaller changes. If the peer's genre also matches the current genre and both mood axes are within threshold, the curation is skipped --- the peer signal confirms alignment without triggering action.

\paragraph{Genre affecting.} For same-domain peers (two MeloTune instances), genre is exchanged alongside mood through a structured \texttt{genre:} prefix in the CMB's \emph{focus} field. The receiver extracts the peer's genre and passes it to the curation pipeline, so co-listening peers converge on both mood and genre. A separate genre-change cooldown (5~minutes) prevents genre ping-pong. Non-music agents (e.g.\ Claude Code) influence mood only, never genre --- they have no genre authority.

\subsection{Personal Arousal Function}
\label{sec:paf}

\subsubsection{The gap: audio intensity is not psychological arousal}

The Russell circumplex defines arousal as a property of the \emph{listener's} affect state --- the degree of psychological activation experienced during listening. The audio-feature APIs of major music catalogs expose a related but fundamentally different quantity: the \emph{recording's} acoustic intensity, derived from spectral energy, loudness, tempo, and onset rate. MeloTune's initial pipeline, following standard practice in music emotion recognition \citep{aljanaki2017developing}, treats audio intensity as a direct proxy for listener arousal via a linear mapping. This conflation produces systematic errors in at least three well-defined scenarios: (1)~\textbf{Habituation} --- a listener who has heard a high-intensity track hundreds of times may experience low arousal from it; (2)~\textbf{Genre-specific affect} --- a funeral march (low-to-moderate audio intensity) reliably produces high arousal because grief is activating, and horror scores achieve high tension through sparse, quiet textures; (3)~\textbf{Novelty response} --- a listener's first encounter with an unfamiliar genre produces higher arousal than repeat exposure to the same audio intensity.

No published system we are aware of models this distinction. Collaborative filtering converges to the population mean; it does not diverge from it per listener.

\subsubsection{Decomposition}

We introduce a \textbf{Personal Arousal Function (PAF)} that decomposes predicted arousal into a population-level prior and a learned per-user adjustment:
\begin{equation}
a(t, u) = \text{prior}(t) + \delta(g_t, \tau, u) \cdot c(g_t, \tau, u)
\label{eq:paf}
\end{equation}
where $t$ is the track, $u$ is the listener, $g_t$ is the track's genre cluster, $\tau$ is the time-of-day band, $\text{prior}(t)$ is the MEI classification (audio-derived, static per track --- the population average), $\delta$ is the learned arousal adjustment (bounded to $[-0.5, +0.5]$), and $c \in [0, 1]$ is a confidence scalar that gates the adjustment magnitude based on sample count. The result is clamped to $[-1, +1]$ on the Russell arousal axis. When no behavioral data exists for a listener--genre--time combination, $\delta = 0$ and $c = 0$, producing the unmodified prior. This zero-impact default ensures PAF cannot degrade the baseline for new users or unfamiliar genres.

\subsubsection{Behavioral signals as training data}

PAF learns $\delta$ from implicit behavioral signals observed during playback. Each signal carries a signed arousal adjustment indicating whether the MEI prior over- or under-estimated the listener's actual arousal response: skip within 15\,s (strong mismatch), skip at 15--60\,s (partial mismatch), completion (acceptable match), completion with favorite (strong positive match), repeat play (reliable affect producer), volume increase (seeking more intensity), and volume decrease (seeking less intensity). When a user-declared mood is available (via the mood dial --- the UEA layer of the EI-3 architecture), skip signals are \textbf{directionally conditioned}: if the declared arousal exceeds the MEI prior, a skip indicates under-arousal; if the declared arousal is below the prior, a skip indicates over-arousal. This directional conditioning ensures the learning signal is correct rather than directionally ambiguous.

\subsubsection{Learning algorithm}

PAF maintains a \textbf{Personal Arousal Profile} consisting of arousal adjustments bucketed by genre cluster and time-of-day band. Each bucket is updated via an exponential moving average (EMA):
\begin{equation}
\delta_{n+1} = \alpha \cdot s_n + (1 - \alpha) \cdot \delta_n
\end{equation}
where $s_n$ is the signed arousal signal from the $n$-th behavioral observation and $\alpha = 0.15$ is the learning rate. The conservative $\alpha$ produces a half-life of approximately 4~sessions. Confidence scales linearly with sample count: $c = \min(1, n / n_{\text{full}})$ where $n_{\text{full}} = 20$. This means 2--3~sessions of a genre are required before PAF reaches full effect.

\subsubsection{Drift as a learning signal}

The EI-3 architecture monitors drift between the user's declared mood (UEA) and the machine's inference (MEI). PAF reinterprets this drift as a \textbf{training signal}. Consistent positive arousal drift for a genre (the user repeatedly declares higher arousal than MEI predicts for classical music) means this listener finds that genre more activating than the population average. PAF integrates genre-contextualised drift entries into the EMA alongside behavioral signals, providing a second, complementary learning channel. This reinterpretation preserves the clarifier dialog for extreme drift while using moderate, consistent drift as evidence of personal arousal deviation.

\subsubsection{Integration boundary}

PAF applies at a single point in the prediction pipeline: after \texttt{MEIEngine} produces a population-level classification and before the result enters the curation pipeline. All downstream consumers --- ERE, the proactive curator, the curation pipeline, the mesh broadcast --- receive the personalised arousal without modification. The CfC trajectory model at Layer~6 predicts a mood trajectory over personalised arousal values rather than population averages, which means the trajectory itself becomes personalised: two listeners hearing the same sequence of tracks follow different predicted trajectories because PAF adjusts the arousal input differently for each.

Once sufficient behavioral data accumulates ($c = 1.0$ in at least one genre bucket), PAF triggers a batch re-classification of the user's existing favorites library, retroactively applying personalised arousal adjustments to previously classified tracks. The ``same track, different prediction for different listeners'' property applies to the full library, not only to tracks encountered after PAF reaches confidence. The re-classification is debounced by a minimum sample threshold (20~new behavioral signals between passes) to avoid redundant batch processing.

\section{Training}
\label{sec:training}

This section documents the training procedure for the \textbf{listener-level CfC} described in \S\ref{sec:cfc-dynamics} --- the model that maintains a private trajectory for one listener and drives that listener's proactive curation. The model is trained offline in PyTorch, exported to CoreML, and shipped in the iOS application bundle. There is no on-device fine-tuning. The mesh-runtime CfC at Layer~6 of MMP (\S\ref{sec:mesh-substrate}) is a \textbf{separate model with its own training regime, documented in the SVAF/MMP paper} \citep{xu2026svaf}, and is not the subject of this section. Per-epoch curves and ablation studies for the listener-level CfC are reported in the extended companion paper.

\subsection{Data}
\label{sec:training-data}

Training sequences are drawn from logged \textbf{Cognitive Memory Block (CMB)} sessions captured by the MeloTune client during first-author usage. The training corpus comprises 204~sessions containing 872~events, spanning December 2025 -- January 2026. Each CMB encodes a listening event with its CAT7 cognitive fields, the listener's affective state at that instant, and contextual annotations (time-of-day, session intent, mesh trigger source). Sessions are chunked into sequences of length $L \in [5, 100]$. Inter-event $\Delta_k$ are preserved exactly so that the closed-form cell update receives the wall-clock intervals it expects at inference.

We apply two augmentations during training: \textbf{time warping}, which rescales $\Delta_k$ within bounded ratios so that the model becomes robust to global tempo changes in event arrival, and \textbf{input noise injection}, which perturbs the per-event input vector to discourage overfitting on the relatively small number of distinct sequences.

\subsection{Loss}
\label{sec:loss}

The training objective is a weighted sum of four terms,
\begin{equation}
\mathcal{L}
= w_T\,\mathcal{L}_\text{traj}
+ w_P\,\mathcal{L}_\text{pat}
+ w_I\,\mathcal{L}_\text{int}
+ w_F\,\mathcal{L}_\text{pred},
\end{equation}
with
\begin{itemize}
\item $\mathcal{L}_\text{traj}$ --- mean squared error on the trajectory head's six outputs against per-event ground truth (emotion, energy, velocities, stability, confidence). $w_T = 1.0$.
\item $\mathcal{L}_\text{pat}$ --- binary cross-entropy on the nine pattern-head sigmoid outputs against pattern labels assigned by an unsupervised clusterer over historical sessions. $w_P = 0.5$.
\item $\mathcal{L}_\text{int}$ --- categorical cross-entropy on the intent head over six coarse session-intent classes. $w_I = 0.5$.
\item $\mathcal{L}_\text{pred}$ --- mean squared error on the next-event predictions of emotion and energy, encouraging the model to produce a useful one-step-ahead forecast. $w_F = 0.3$.
\end{itemize}

The trajectory term is the headline loss: it is the signal that the proactive curation policy ultimately consumes.

\subsection{Optimisation}
\label{sec:optimisation}

We use AdamW with learning rate $10^{-3}$ and a cosine annealing schedule. Gradients are clipped at unit norm. Validation loss is monitored at the end of each epoch with early stopping after a patience of 10~epochs without improvement, and checkpoints are saved every 10~epochs. The model is exported through a flattened-output wrapper (\texttt{LNNForCoreML}) for deployment on iOS~15.0+.

\subsection{Offline performance}
\label{sec:offline-performance}

The final model (94,552~parameters, early-stopped at epoch~23) achieves the following on the held-out validation split (15\% of sessions): trajectory MAE of $0.414$ (emotion MAE $11.9$, energy MAE $18.2$ on the $[0, 99]$ scale), pattern head accuracy $96.6\%$, and intent head accuracy $69.4\%$. The high pattern accuracy reflects that temporal patterns (morning routine, evening wind-down) are strongly time-correlated. The lower intent accuracy reflects the inherent ambiguity of intent labels in the absence of explicit user input --- a listener's intent is inferred from behavioral signals that are consistent with multiple plausible intents.

\subsection{What is not learned on-device}
\label{sec:not-learned}

The CfC model weights are \textbf{frozen} after deployment. Hidden state is persisted locally to the device's user defaults across application launches with a 24-hour validity window, but the network weights themselves are not updated. This is a deliberate privacy choice: no listening history, no gradient signal, and no per-user training data leaves the device, and no per-user model is maintained on a server.

Per-listener adaptation is provided by the Personal Arousal Function (\S\ref{sec:paf}), which learns arousal adjustments from behavioral signals at the classification boundary without gradient computation. The CfC hidden state provides within-session temporal context; PAF provides across-session personalisation. Together, the frozen CfC and the adaptive PAF span both timescales without requiring on-device gradient updates. We discuss the implications in \S\ref{sec:discussion}.

\section{Evaluation Protocol and Qualitative Observations}
\label{sec:evaluation}

This paper reports the \textbf{architecture} and \textbf{deployment} of MeloTune together with \textbf{qualitative observations} from the live system. A controlled deployment evaluation, with quantitative comparisons against baselines, is the subject of an extended companion paper currently in preparation. We describe the planned evaluation here in full so that the present results can be interpreted in context, and so that the protocol itself can be reviewed independently.

\subsection{What this paper reports}
\label{sec:what-reported}

We report (i)~the deployed architecture in full detail (\S\ref{sec:method}), (ii)~the training procedure and offline performance of the CfC trajectory head on logged CMB sequences (\S\ref{sec:training}), and (iii)~qualitative observations from the live App Store deployment, including representative session traces and the curation policy's behaviour under typical usage patterns. We do not report next-item ranking metrics, controlled user studies, or quantitative head-to-head comparisons against sequential-recommendation baselines in this version. The remainder of \S\ref{sec:evaluation} specifies the protocol under which those measurements are being collected.

\subsection{Research questions}
\label{sec:research-questions}

The extended evaluation is structured around four research questions.

\textbf{RQ1 --- Trajectory accuracy.} How accurately does the CfC predict the listener's affective trajectory over a planning horizon of 5~minutes, relative to the user-supplied mood-meter ground truth and to standard sequential-recommender baselines that produce no trajectory at all?

\textbf{RQ2 --- Proactive vs reactive curation.} Does proactive trajectory-driven curation reduce skip rate, increase session length, and improve listener-reported satisfaction relative to a reactive ablation in which the same code path queues for the \emph{current} mood rather than the \emph{projected} mood?

\textbf{RQ3 --- Generalisation across listeners.} Does a single shipped CfC model generalise across listeners without per-user fine-tuning, or does the architecture require on-device adaptation to perform acceptably for individual listening profiles?

\textbf{RQ4 --- Co-listening coherence at the mesh-runtime CfC.} When two devices participate in a mesh-coupled session, does the \textbf{shared mood field} maintained by the per-agent Layer-6 CfC inside the SYMCore mesh runtime converge on a coherent attractor whose coherence signal $\rho(t)$ exceeds an independent-listening control, and when the listener-level proactive curator subscribes to that signal, does the resulting biased curation produce subjectively shared listening experiences? Note that this RQ is asked at the \textbf{mesh-runtime CfC}, not at the listener-level CfC; the listeners' private CfC hidden states are never compared, because they never leave their respective devices. Coherence is measured against the \emph{public} CMB-derived shared mood field, not against private hidden states.

\textbf{RQ5 --- PAF divergence.} Does the Personal Arousal Function produce meaningfully different arousal predictions for different listeners on the same track, and does the PAF-adjusted arousal prediction reduce skip rate compared to the population-level MEI prior? Preliminary single-listener evidence is reported in \S\ref{sec:paf-evidence}; multi-listener measurement is planned for the extended evaluation.

\subsection{Data collection}
\label{sec:data-collection}

\textbf{Source.} All evaluation data is collected from the live MeloTune application via an opt-in research-logging facility (\texttt{ResearchLogger}). Logging is gated behind an explicit consent screen presented on first launch of the research build; users may revoke consent at any time and delete their logged sessions from the device.

\textbf{Sessions.} A session is defined as a contiguous period of music playback bounded by an application launch event and either an exit, a backgrounding longer than ten minutes, or an explicit session-end event. Each session yields a structured event log.

\textbf{Per-event fields.} For each event within a session we record: timestamp, event type (play / skip / pause / resume / mood-meter update / curation tick / peer state update), the playing track's catalog identifier and audio-feature vector, the user-supplied mood-meter input if any, the CfC's predicted trajectory at that instant (current point, projected point at the planning horizon, velocity, stability, confidence), the curation policy's chosen action and its reasoning, and the set of currently-connected mesh peers with their last-known hidden-state divergence and coupling weights.

\textbf{Personal sessions.} We target ${\sim}200$ personal sessions across ${\sim}10$ listeners, with sessions of typical length (15--90~minutes). Listeners include the first author and TestFlight participants who have explicitly opted into research logging.

\textbf{Co-listening sessions.} We target ${\sim}30$ co-listening sessions. Each co-listening session involves two devices simultaneously running the research build, joined to the same mesh, listening through the same session timeline. Co-listening sessions are run on the first author's own hardware (an iPhone and a Mac Catalyst instance) and on TestFlight pairs that opt in.

\subsection{Baselines}
\label{sec:baselines}

We compare against three baselines.

\textbf{Reactive ablation.} The same MeloTune code path with the CfC's predictive head bypassed: the curation policy receives the \emph{current} predicted mood instead of the projected mood at the planning horizon. This isolates the contribution of forward projection from every other architectural choice. It is the cheapest baseline and the most diagnostic.

\textbf{GRU4Rec on logged sequences.} A standard session-based recurrent recommender \citep{hidasi2016session} trained on the same listening sequences harvested from the research log. This is the canonical sequential-recsys baseline that reviewers will expect.

\textbf{Random-within-genre.} A sanity baseline that draws tracks uniformly from the genres associated with the current 400-anchor mood lookup cell. Distinguishes architectural contribution from naive mood-conditioned retrieval.

A qualitative comparison against the platform's own black-box ``radio'' feature is reported separately and is not used to make controlled claims; it is included only to situate MeloTune's behaviour in the experience that listeners actually have.

\subsection{Metrics}
\label{sec:metrics}

\textbf{Trajectory MSE (RQ1).} Mean squared error between the CfC's predicted trajectory at horizon $\tau_p$ and the user-supplied mood-meter ground truth at the corresponding wall-clock time. Reported per-axis (valence, arousal) and as a Euclidean trajectory error.

\textbf{Skip rate (RQ2).} Fraction of queued tracks skipped before some fraction of their length is played, computed per session and aggregated across the user population. Reported separately for the proactive and reactive arms.

\textbf{Session length and time-to-first-skip (RQ2).} Median and distribution of session length and time-to-first-skip, as proxies for sustained listener engagement.

\textbf{Listener-reported satisfaction (RQ2).} A short post-session survey, presented in the application, asking listeners to rate the session on five-point scales for \emph{felt understood}, \emph{would listen again}, and \emph{fit the moment}. Reported as Likert distributions.

\textbf{Cross-listener generalisation (RQ3).} Held-out per-listener evaluation: for each listener, compute trajectory MSE and skip rate against a model trained on the \emph{other} listeners only. We do not fine-tune per user.

\textbf{Mesh coherence (RQ4).} For each co-listening session, the time-series of the mesh-runtime CfC's published coherence signal $\rho(t)$, the fraction of session time during which incoming peer CMBs are admitted in each SVAF band (aligned / guarded / rejected / redundant), and the time-to-convergence on a shared mood field. Compared against an independent-listening control in which the same two listeners listen on the same hardware with mesh participation disabled. \textbf{Pairwise listener-CfC hidden states are not compared}, because under MMP they never leave their respective devices.

\subsection{Splits and statistical protocol}
\label{sec:splits}

Sessions are partitioned chronologically: the first 80\% by timestamp form the training split, the remaining 20\% the held-out test split. This avoids the look-ahead bias that arises when sessions from the same listener are split randomly. Significance is reported at $p < 0.05$ using paired bootstrap resampling on session-level metrics.

\subsection{Qualitative observations from the live deployment}
\label{sec:qualitative}

In the absence of the controlled measurements above, we report four qualitative findings from the first author's own use and from logged TestFlight sessions.

\textbf{Observation 1 --- Trajectory anticipation.} In typical evening listening sessions the CfC's projected mood at horizon $\tau_p \approx 5$\,min frequently anticipates a downshift in arousal that the user-supplied mood meter then confirms several minutes later. The proactive curator re-queues into lower-arousal material before the listener has explicitly requested it. This is the architecture working as designed; the magnitude and frequency of this effect across listeners is the subject of RQ1 and RQ2.

\textbf{Observation 2 --- Stability of confidence-gated curation.} The $\text{confidence} \geq 0.4$ gate on the curation policy is in practice rarely a binding constraint during established sessions and is most often binding in the first minutes after a cold start, when the CfC has not yet had enough events to converge on a confident trajectory. The policy degrades gracefully to no-action under low confidence, which is the intended behaviour.

\textbf{Observation 3 --- Mesh substrate behaviour under co-listening.} When two devices on the same SYM mesh listen together, the SVAF Layer-4 evaluator reliably begins admitting incoming peer CMBs in the \emph{aligned} band ($\delta_\text{total} \leq 0.25$) within the first few minutes of joint listening, with the \emph{mood} field admitted across virtually every CMB (consistent with the protocol's R5 guarantee that mood always crosses domain boundaries). Mood affecting is verified end-to-end: a mood change on Device~A triggers curation on Device~B within 10~seconds. Genre affecting is also verified: when Device~A switches genre, Device~B curates the same genre via the structured \texttt{genre:} prefix in the CMB \emph{focus} field. Cross-platform mesh verified: iOS $\leftrightarrow$ macOS Catalyst $\leftrightarrow$ Windows $\leftrightarrow$ Node.js (April 2026).

\textbf{Observation 4 --- Echo loop prevention.} During two-device testing, an echo loop was observed and fixed: two MeloTune instances ping-ponged mood changes (A broadcasts $\to$ B curates $\to$ B broadcasts $\to$ A curates $\to$ loop). The root cause was that the Emotional Resolution Engine (ERE) could not distinguish organic mood (from user actions) from mesh-induced mood (from curation triggered by a peer's CMB). The fix --- ERE isolation with a 60-second window (MMP~\S8.2) plus lineage-based detection (\S15.2) --- eliminates the loop in all observed scenarios. This finding was sufficiently general to warrant formalisation in the protocol specification as the organic mood constraint.

\subsection{Threats to validity (planned evaluation)}
\label{sec:threats}

We list these in advance because they shape the protocol above.

\textbf{Population.} The opt-in TestFlight cohort skews toward early adopters and may not represent the full distribution of music listeners. We acknowledge this and do not claim population-level generalisation from the first round of measurements.

\textbf{Mood-meter as ground truth.} Treating the user's explicit mood input as ground truth conflates two sources of error: the CfC's prediction error and the user's introspective error. We use this conflation deliberately, on the grounds that the user's reported state is the only operationally meaningful target for a music recommender, but we report it as a limitation rather than a solved problem.

\textbf{Catalog effects.} Apple Music and Spotify expose different audio feature ranges and different editorial structures. Where possible we report metrics conditioned on catalog. Cross-catalog claims are marked explicitly.

\textbf{Self-reporting bias.} The post-session satisfaction survey is voluntary and may suffer from selection effects. We report response rates alongside results.

\begin{figure*}[t]
\centering
\includegraphics[height=0.45\textheight]{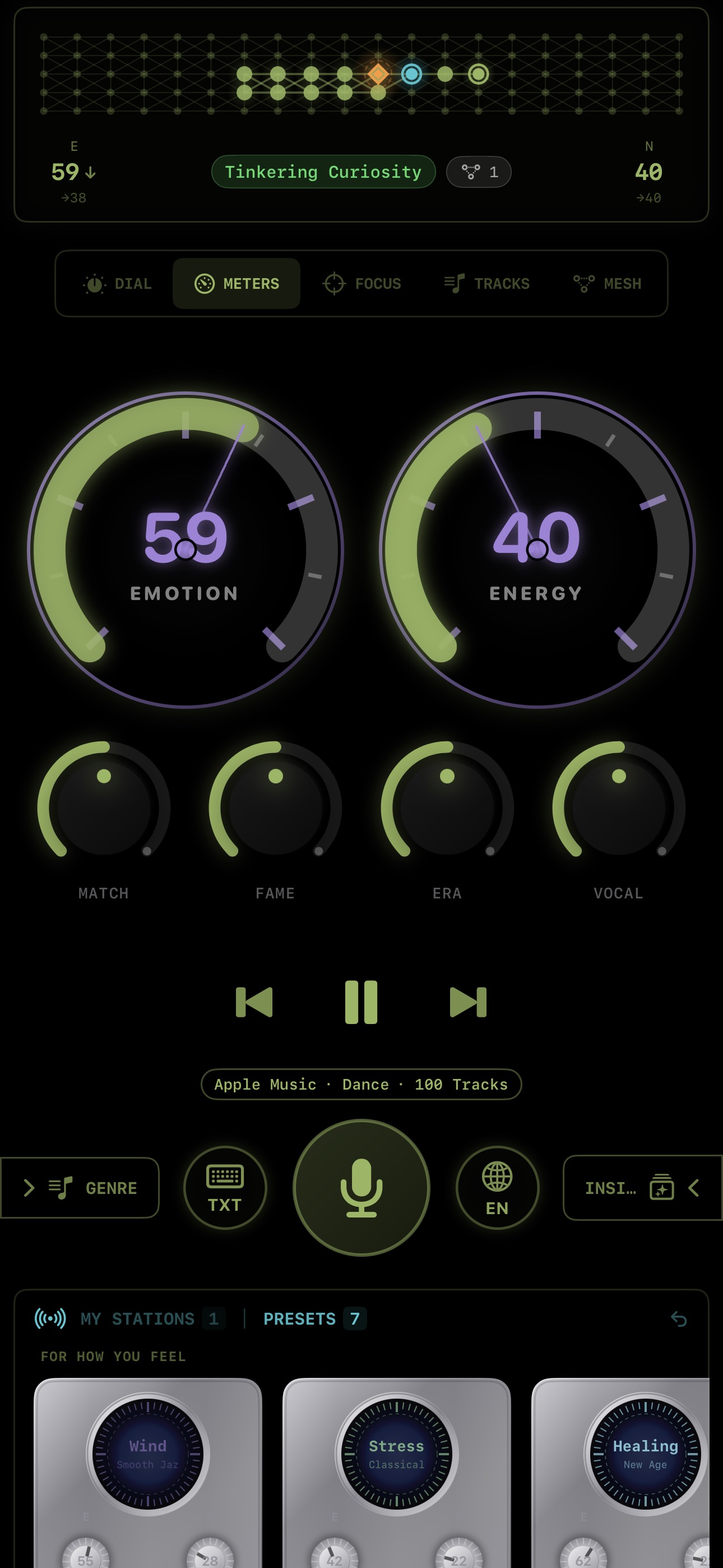}\hspace{1pt}
\includegraphics[height=0.45\textheight]{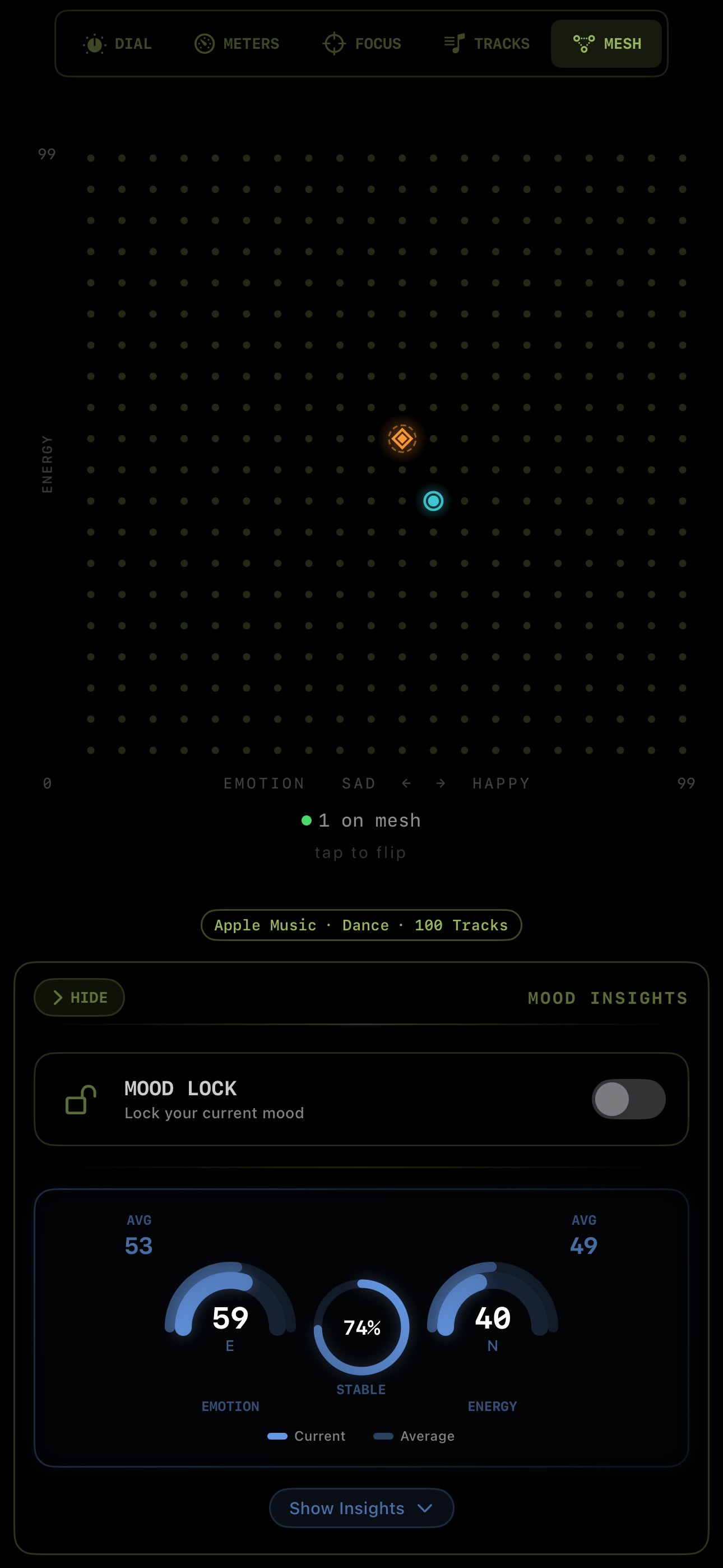}\hspace{1pt}
\includegraphics[height=0.45\textheight]{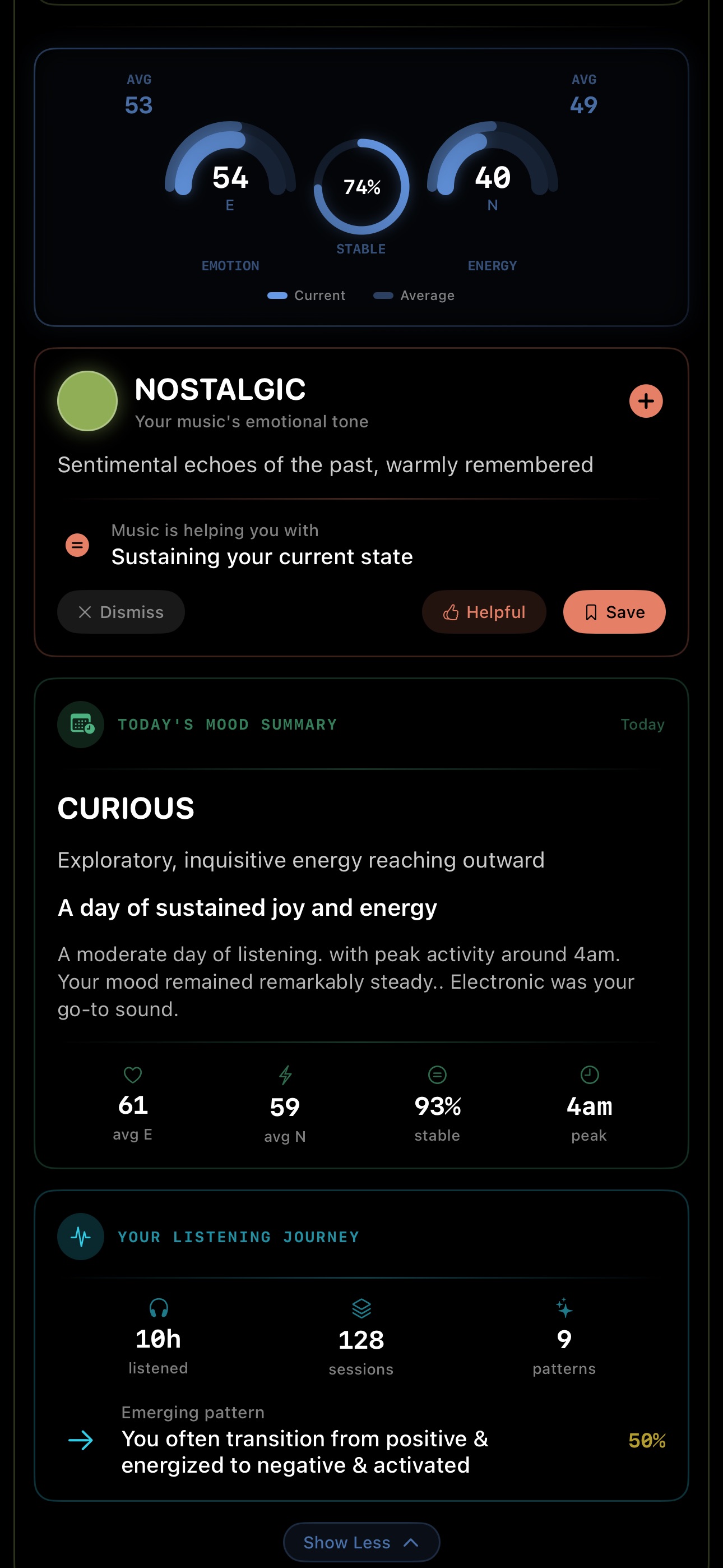}
\caption{MeloTune deployed interface. \textbf{Left:} Meters mode showing Emotion (59) and Energy (40) dials with filter knobs (Match, Fame, Era, Vocal), station presets, and voice/text input. \textbf{Centre:} Mesh mode showing the Russell circumplex grid with the local user (cyan) and a connected peer (orange diamond), ``1 on mesh'' indicator, and mood insights panel. \textbf{Right:} Insights panel showing current mood state (Nostalgic), daily mood summary (Curious), listening journey statistics (10h, 128 sessions, 9 patterns), and detected behavioral pattern.}
\label{fig:ui-screenshots}
\end{figure*}

\subsection{PAF: live deployment evidence}
\label{sec:paf-evidence}

To validate the PAF learning loop described in \S\ref{sec:paf}, we present
behavioral data captured from a live deployment session (12~April~2026,
afternoon--evening, single listener, Apple Music). The session produced
46~behavioral observations across 11~genres and 2~time-of-day bands, capturing
all five signal types: completion, skip-early ($<15$\,s), skip-mid
($15$--$60$\,s), volume-up, and volume-down.
Table~\ref{tab:paf-evidence} lists 18~representative signals (full log in
supplementary material). Table~\ref{tab:paf-profile} shows the resulting
PAF profile --- 12~genre$\times$time arousal adjustments learned from these
observations.

\begin{table*}[t]
\centering
\caption{Representative behavioral responses from a live deployment session (12 April 2026, afternoon--evening, single listener).
MEI\textsubscript{a} = population-level audio-derived arousal prior;
UEA\textsubscript{a} = user-declared arousal;
$\Delta$ = MEI\textsubscript{a} $-$ UEA\textsubscript{a}.
46~total observations across 11~genres; 18~representative entries shown (full log in Appendix~\ref{sec:appendix-paf}).}
\label{tab:paf-evidence}
\begin{tabular}{rllllrrrr}
\toprule
\# & Genre & Time & Signal & Pos & MEI\textsubscript{a} & UEA\textsubscript{a} & $\Delta$ \\
\midrule
1 & pop & aft & skipMid & 0 & $+0.40$ & $+0.10$ & $+0.30$ \\
2 & pop & aft & skipEarly & 2 & $-0.28$ & $+0.10$ & $-0.38$ \\
3 & jazz & aft & completed & 0 & $-0.48$ & $+0.00$ & $-0.48$ \\
4 & jazz & aft & skipMid & 0 & $-0.48$ & $-0.52$ & $+0.04$ \\
5 & house & aft & completed & 0 & $+0.04$ & $-0.01$ & $+0.05$ \\
6 & dance & aft & skipMid & 3 & $-0.16$ & $-0.01$ & $-0.15$ \\
9 & hard\_rock & aft & skipEarly & 0 & $-0.04$ & $+0.00$ & $-0.04$ \\
12 & hard\_rock & aft & completed & 0 & $+0.76$ & $+0.76$ & $+0.00$ \\
14 & classical & aft & completed & 0 & $-0.30$ & $+0.00$ & $-0.30$ \\
16 & electronic & aft & skipMid & 3 & $+0.22$ & $+0.00$ & $+0.22$ \\
35 & electronic & eve & completed & 0 & $-0.40$ & $+0.10$ & $-0.50$ \\
37 & electronic & eve & completed & 1 & $+0.44$ & $+0.10$ & $+0.34$ \\
39 & electronic & eve & volumeDown & 4 & $-0.10$ & $+0.10$ & $-0.20$ \\
40 & electronic & eve & volumeUp & 4 & $-0.10$ & $+0.10$ & $-0.20$ \\
41 & Cantopop & eve & completed & 0 & $-0.30$ & $+0.00$ & $-0.30$ \\
42 & Mandopop & eve & volumeUp & 5 & $-0.30$ & $+0.00$ & $-0.30$ \\
45 & dance & eve & completed & 0 & $+0.04$ & $+0.18$ & $-0.14$ \\
46 & edm & eve & completed & 0 & $+0.70$ & $+0.42$ & $+0.28$ \\
\bottomrule
\end{tabular}
\end{table*}

\begin{table}[t]
\centering
\caption{Learned PAF profile after 46 behavioral observations across 11~genres and 2~time-of-day bands.
$\Delta_a$ = EMA-smoothed arousal adjustment ($\alpha{=}0.15$);
\textit{conf} = confidence (sample count\,/\,20).
Pop reaches full confidence ($c{=}1.0$) after 22~observations.}
\label{tab:paf-profile}
\begin{tabular}{llrrr}
\toprule
Genre & Time & $\Delta_a$ & Conf & $n$ \\
\midrule
cantopop & eve & $+0.0239$ & 0.05 & 1 \\
classic\_rock & aft & $+0.0481$ & 0.05 & 1 \\
classical & aft & $+0.0499$ & 0.10 & 2 \\
dance & aft & $+0.1451$ & 0.05 & 1 \\
dance & eve & $+0.0075$ & 0.05 & 1 \\
electronic & aft & $-0.1420$ & 0.05 & 1 \\
electronic & eve & $+0.0311$ & 0.30 & 6 \\
hard\_rock & aft & $-0.1107$ & 0.10 & 2 \\
house & aft & $+0.0484$ & 0.05 & 1 \\
jazz & aft & $+0.0047$ & 0.20 & 4 \\
mandopop & eve & $+0.0191$ & 0.15 & 3 \\
pop & aft & $-0.1788$ & 1.00 & 22 \\
\bottomrule
\end{tabular}
\end{table}

Four patterns are visible in the data. First, \textbf{directional skip
conditioning} operates correctly: pop entry~1 ($\Delta = +0.30$, skipped)
indicates the MEI prior over-estimated arousal relative to the listener's
declared state, driving PAF to accumulate a strong negative adjustment for pop
($\Delta_a = -0.179$). Second, \textbf{confidence gating reaches saturation}:
pop accumulates 22~observations reaching full confidence ($c = 1.0$), at which
point the $-0.179$ adjustment applies at full strength --- the system
committed to a strong arousal reduction only after sufficient evidence. Third,
\textbf{time-of-day differentiation} emerges naturally: electronic music in the
afternoon ($\Delta_a = -0.142$, skipped) diverges from electronic in the
evening ($\Delta_a = +0.031$, mostly completed), reflecting the same listener's
different arousal relationship to the same genre at different times. Fourth,
\textbf{volume signals} are captured alongside skip/completion signals
(entries~39--44), providing a complementary engagement channel that does not
require the listener to terminate the track.

These results are from a single listener and do not constitute a controlled
evaluation. They demonstrate that the PAF learning loop is operational
end-to-end across all five signal types, that directional conditioning is
applied correctly, that EMA updates accumulate to full confidence, and that the
resulting 12-bucket profile is structurally consistent with the listener's
observed preferences across genres and time of day. Quantitative evaluation
with multiple listeners and session-level metrics is reported in the companion
paper.

\section{Discussion}
\label{sec:discussion}

We discuss the architectural choices that distinguish MeloTune from sequential music recommenders, the implications of an on-device deployment, the limitations we are aware of, and the implications for industrial recommender systems.

\subsection{Why continuous time}
\label{sec:why-continuous}

The choice of a closed-form continuous-time network is not incidental. A standard recurrent or transformer recommender that is trained on logged listening sequences must commit, at training time, to a fixed step rate or to a positional encoding that is at best a weak proxy for wall-clock time. Real listening is bursty: a listener may skip three tracks in twenty seconds and then play a single song for the next eight minutes. A discrete-step model treats both intervals as one step. A continuous-time model treats them as 20~seconds and 8~minutes, which is what they are.

The closed-form variant is what makes the choice deployable. A neural ODE with an adaptive solver inside the inner loop of an iPhone application is feasible in principle but adds variance to inference latency that is not acceptable in user-facing real-time code. The analytic update of the CfC is single-pass, deterministic in latency, and completes in under a millisecond on the hardware we ship to.

\subsection{Why Russell's circumplex}
\label{sec:why-russell}

We use the two-axis valence/arousal plane because it matches the intrinsic dimensions exposed by the audio-feature APIs of the major music catalogs. Apple Music's energy and valence attributes, Spotify's named features of the same form, and the dominant representations in the MER literature are all variants of the same two-dimensional construct. A higher-dimensional affect space (e.g.\ adding dominance and tension) is psychologically richer but offers no operational advantage when the catalog itself is two-dimensional. We keep the model simple at the boundary the catalog provides.

The 400-anchor lookup that discretises the plane is a deliberate intermediate representation. It gives the user interface a stable vocabulary of named moods, gives the retrieval head a deterministic indexing scheme, and gives the trajectory head a quantised target space against which generalisation can be evaluated. The discretisation is authored, not learned, which makes it inspectable and editable without retraining.

\subsection{Why on-device}
\label{sec:why-on-device}

On-device inference is not a deployment convenience --- it is a protocol requirement. MMP \citep{xu2026mmp} defines each node as a sovereign cognitive entity with private Layer-6 state. Server-side inference would require hidden-state transmission to a central node, violating guarantee R4 (hidden states never cross the wire). The only deployment mode consistent with the protocol is on-device.

Beyond protocol alignment, on-device is required by the learning architecture. The Personal Arousal Function (\S\ref{sec:paf}) learns from behavioral signals --- skip timing, volume changes, session position, completion ratio --- that are observable only at the playback device. Server-side PAF would require a real-time behavioral stream sent per listening event, adding latency, privacy exposure, and server dependency to every track transition. On-device PAF observes these signals at zero cost and zero latency.

The frozen-model limitation often cited against on-device ML --- that adaptation is constrained to what the hidden state can encode --- is partially resolved by PAF itself. PAF provides on-device adaptation that does not require gradient computation or model retraining: the EMA-based arousal adjustments accumulate from behavioral signals, reach full confidence after ${\sim}20$ observations per genre bucket, and persist across sessions. This is not parameter adaptation in the traditional sense, but it is personalisation that a frozen server-side model serving multiple listeners cannot provide without per-user state.

More broadly, on-device enables \textbf{agent-native learning}: each agent learns independently from its own experience, broadcasts conclusions as CMBs, and receivers decide what to absorb through SVAF per-field drift evaluation. This is fundamentally different from federated learning, where a central server aggregates gradients. In the mesh model, there is no central server and no gradient. Each agent is sovereign; the mesh is emergent.

\subsection{Why two CfCs}
\label{sec:why-two-cfcs}

Collective intelligence requires separating what an agent knows privately from what the group knows collectively. The CMB is the boundary between them. The two-CfC architecture is the concrete realisation of this separation --- private cognition for the individual agent, shared cognition for the mesh, with typed semantic fields as the only interface between the two.

The listener-level CfC described in \S\ref{sec:cfc-dynamics} is private to one listener: it sees that listener's event stream, maintains that listener's affective trajectory, and never broadcasts its hidden state. The mesh-runtime Layer-6 CfC inside SYMCore (\S\ref{sec:mesh-substrate}) is shared at the mesh level: it sees structured CMBs broadcast from peers, integrates them through SVAF Layer-4 evaluation, and produces a coherence signal for the room. The two CfCs are independent --- different weights, different training regimes, different latent spaces, different roles. They communicate only through CMBs at the event level.

Four properties follow from this separation. First, \textbf{privacy by construction}. CfC hidden states encode the finest-grained information the system has about an individual listener; broadcasting them would either leak that information or require cross-device alignment machinery (federated training, secure aggregation) that does not fit on a phone in real time. Restricting inter-device communication to structured CMBs with explicitly typed, per-field admissible fields puts a hard upper bound on what one listener can ever learn about another.

Second, \textbf{modularity}. A future improvement to the listener-level CfC does not touch the mesh runtime, and a future improvement to the mesh runtime does not require any listener model to be retrained. The CMB schema is the public interface; everything inside each CfC is private.

Third, \textbf{principled echo-loop prevention}. When Device~B receives a peer CMB and curates a new playlist in response, the track mood of the new playlist is a mesh-induced signal, not an organic change in the listener's internal state. If ERE were to fuse this track mood and broadcast it as B's own mood, the result is an echo loop: A broadcasts $\to$ B curates $\to$ B's ERE absorbs the new track mood $\to$ B broadcasts $\to$ A curates $\to$ repeat. The two-CfC separation resolves this by construction. The listener-level CfC tracks only the listener's organic trajectory --- derived from non-mesh inputs (continued listening, explicit user action, temporal decay). Mesh-curated track mood is excluded from ERE fusion for a configurable isolation window (60~seconds in the shipping implementation). Only after the listener has independently confirmed the mood shift through sustained listening does ERE resume organic fusion and broadcast. This is not a patch on a single-CfC design; it is a property that emerges from the separation itself.

Fourth, \textbf{agent-native learning at each layer}. The listener-level CfC learns from private behavioral signals via PAF; the mesh-runtime CfC learns from peer CMBs via SVAF coupling. Neither learning process contaminates the other. A listener who skips pop tracks (PAF signal) does not cause their peer's mesh-runtime CfC to downweight pop --- only their own listener-level predictions change. The mesh carries conclusions, not training data.

\subsection{Why a mesh substrate}
\label{sec:why-mesh}

Co-listening is not currently a first-class construct in any mainstream music recommender. A car with two listeners gets one listener's queue. A kitchen with a household gets the recently-active account's queue. A workout class gets the instructor's queue. Treating each of these as the union of individual histories misses the \emph{emergent} affect of a shared session.

We use a peer-to-peer mesh rather than a server-mediated co-session because the unit of co-listening is the \emph{physical} room, not a server-side identity. Two devices in the same room form a session without authentication beyond explicit consent on each device. The SVAF coupling layer ensures that two listeners whose hidden states are too far apart simply do not couple --- the rejected regime is a no-op, not a forced compromise. We believe this is the right default for a co-listening system: agreement is opt-in at the representation level, not enforced from above.

\subsection{Limitations}
\label{sec:limitations}

We list limitations grouped by where in the architecture they arise.

\textbf{Encoder.} The current track-level affect inferencer consumes metadata only. Behavioral signals --- skip events, mood-meter inputs, volume changes, session position --- are consumed by PAF (\S\ref{sec:paf}) at the classification boundary, but not yet by the CfC input encoder. A history-aware encoder that incorporates these signals into the CfC's input vector is straightforward but not yet shipped. Until it is, behavioral adaptation operates through PAF's arousal adjustment rather than through richer CfC input.

\textbf{Mesh consumer (closed-loop).} The mesh substrate runs end-to-end: each device emits CMBs over MMP, SVAF Layer~4 evaluates incoming peer CMBs per-field, and the per-agent Layer-6 CfC inside SYMCore integrates the admitted fields and publishes a coherence signal. The remaining gap is at the \textbf{listener-level curator}, which does not yet subscribe to the Layer-6 coherence signal. Mesh-biased curation is therefore an architectural property of the system, not yet a measured behaviour. We are explicit about this throughout the paper and report it as such.

\textbf{Frozen model.} Model weights are frozen; there is no on-device gradient-based fine-tuning. However, PAF (\S\ref{sec:paf}) provides a form of non-gradient on-device adaptation: per-listener arousal adjustments accumulate from behavioral signals via EMA, reach full confidence after ${\sim}20$ observations per genre bucket, and persist across sessions. This is weaker than full fine-tuning --- PAF adjusts the arousal axis only, not valence or genre --- but it is meaningful, privacy-preserving, and requires no server. The remaining limitation is that the CfC hidden state cannot encode listener-specific preferences beyond what PAF captures at the classification boundary.

\textbf{Evaluation.} The present paper reports architecture and deployment, not controlled measurements. We have made the planned protocol explicit in \S\ref{sec:evaluation} so that the present paper can be read as a companion to the forthcoming quantitative work rather than as a replacement for it.

\subsection{Implications}
\label{sec:implications}

We close with implications for two communities. Music is the case study; the substrate is the contribution.

\subsubsection{For streaming-platform recommender teams}

Four implications.

First, \textbf{continuous time is a missing layer}, not a competing architecture. A CfC-based affective trajectory model can sit upstream of an existing sequential recommender and feed it a state that the recommender does not currently have. The two are complementary.

Second, \textbf{on-device deployment of continuous-time models is now practical}. The closed-form variant is small enough, fast enough, and predictable enough in latency to live in the inner loop of a mainstream music application on standard mobile hardware.

Third, \textbf{co-listening is an unserved use case} for which the sequence-of-items framing has no good answer. A peer-to-peer substrate that exchanges structured CMBs and integrates them through a per-agent CfC at Layer~6 provides one. We do not claim it is the only one; we claim it is the simplest one we have found that works on real devices today.

Fourth, \textbf{per-listener arousal prediction is an untapped personalisation axis}. Current recommenders treat audio intensity as a proxy for listener arousal, producing identical predictions for every user. PAF demonstrates that even a simple EMA over behavioral signals produces meaningful per-listener divergence after ${\sim}20$ observations per genre. This requires a declared-mood input surface and on-device behavioral observation --- capabilities that streaming platforms could add without architectural upheaval.

\subsubsection{For agent-platform and multi-agent infrastructure teams}

Three further implications, addressed to readers who are building or operating multi-agent systems outside the music domain.

First, \textbf{MMP/SVAF works in production on consumer mobile hardware.} The MeloTune deployment is the first end-to-end reference implementation of an MMP/SVAF agent at production scale on iOS. The wire-level guarantees of the published specification --- CMBs as the unit of inter-agent communication, per-field drift evaluation at SVAF Layer~4, the per-agent CfC at Layer~6, and the protocol commitment that hidden states never cross the wire --- are enforced by the shipping code. The accompanying SDK release (\texttt{sym-swift}~v0.3.78, \texttt{SYMCore}~v0.3.7) deprecates the legacy hidden-state-broadcast path that earlier MMP versions carried, bringing the wire format into strict conformance with the spec. Teams building agents in other domains --- including safety-critical BCI applications, where a third-party Rust implementation (AxonOS) has independently verified MMP interop --- can build against the same SDKs and the same protocol guarantees.

Second, \textbf{the CAT7 schema is a natural fit for cross-domain agent coordination.} MeloTune populates seven typed fields per CMB: focus, issue, intent, motivation, commitment, perspective, and mood. A fitness agent receiving one of MeloTune's CMBs admits the \emph{mood} field for its own purposes (affect crosses domain boundaries by protocol guarantee R5) and suppresses the \emph{focus} field (the specific track is irrelevant to exercise prescription). A coding agent does the inverse --- admits \emph{focus} and \emph{issue}, suppresses \emph{mood}. The same seven-field decomposition serves every domain without protocol changes; what differs across agents is the per-agent SVAF field weights, not the schema.

Third, \textbf{the two-CfC pattern is general beyond music.} The separation of a private per-agent CfC (which encodes that agent's domain expertise) from a shared mesh-runtime CfC (which integrates incoming CMBs from peers) is the architectural pattern that allows agents to participate in collective intelligence without leaking domain-internal state. The same pattern applies to a code agent participating in a developer's IDE without leaking the source tree; to a health agent participating in a household without leaking the patient record; to a research agent participating in a lab without leaking unpublished hypotheses. Music is a particularly clean demonstration because the private dimension (a listener's full listening history) and the shared dimension (the room's emergent mood) are easy to name, but the architectural separation is general.

\subsection{Organic mood as a protocol constraint}
\label{sec:organic-mood-protocol}

The echo-loop problem described in \S\ref{sec:why-two-cfcs} is not specific to MeloTune. Any mesh of agents that (a)~receive peer signals, (b)~act on them by changing local state, and (c)~broadcast their updated local state will produce echo loops unless the broadcast step distinguishes organic state changes from signal-induced ones. This is a general property of coupled dynamical systems with feedback, not an application-specific bug.

We have therefore elevated the constraint to the protocol level. MMP~\S8.2 specifies that the mood field in a CMB represents the broadcasting agent's \emph{inferred organic mood} --- the agent's own affective state as derived from its local, non-mesh inputs. Mood changes induced by received mesh signals must not be attributed to the broadcasting agent's mood until the agent has independently confirmed the shift through non-mesh inputs. MMP~\S15.8 defines the two-layer defense: (1)~a lineage-based echo check at the protocol layer (reject CMBs whose lineage includes the local agent's own prior broadcast), and (2)~the organic-mood constraint at the application layer (ERE isolation window).

This protocol-level constraint has implications beyond music. A fitness agent that receives a peer's fatigue signal and reduces its own exercise intensity should not broadcast the reduced intensity as its user's organic fatigue --- that would create a fatigue cascade across the mesh. A coding agent that receives a peer's ``debugging frustration'' and adjusts its own communication style should not broadcast the adjusted style as its own mood --- that would create a mood contagion loop. In each case, the organic-mood constraint prevents the mesh from amplifying its own signals through feedback. The constraint is simple to state (broadcast only organic mood) and simple to implement (gate broadcast on a post-signal isolation window), but it is load-bearing for any mesh of affect-coupled agents.

\section{Conclusion}
\label{sec:conclusion}

We presented MeloTune, an iPhone-deployed application that instantiates MMP as a music agent and uses the deployment as a case study to demonstrate that the substrate is real, the two-cognition-layer architecture is implementable on consumer mobile hardware, and the protocol guarantees hold in a shipping product. Each device runs two distinct closed-form continuous-time networks: a private listener-level CfC that predicts a short-horizon affective trajectory on Russell's circumplex and drives proactive music curation, and a shared mesh-runtime CfC at MMP Layer~6 that integrates Cognitive Memory Blocks broadcast by co-listening peers and produces a coherence signal for the room. The two CfCs do not share hidden state; they communicate only through CMBs at the event level, evaluated at SVAF Layer~4 under per-field drift bounds. A Personal Arousal Function learns per-listener arousal adjustments from behavioral signals, replacing the linear audio-intensity-to-arousal mapping with a personalised prediction that produces different arousal scores for different listeners on the same track. All inference runs on-device. The accompanying SDK release (\texttt{sym-swift}~v0.3.78, \texttt{SYMCore}~v0.3.7) enforces strict MMP~v0.2.2 conformance: CfC hidden states never cross the wire, the organic mood constraint (\S8.2) prevents echo loops at the protocol level, and Bonjour auto-reconnect maintains mesh presence through iOS backgrounding.

This paper makes five contributions: (1)~a \textbf{Personal Arousal Function} that learns per-listener emotional response from behavioral signals and UEA--MEI drift, replacing the linear audio-intensity-to-arousal mapping with a personalised prediction --- the capability that distinguishes MeloTune from population-mean recommenders; (2)~a \textbf{two-CfC, two-cognition-layer architecture} that separates private per-listener trajectory prediction (with proactive curation) from shared mesh-runtime coupling, deployed on-device via CoreML; (3)~an \textbf{organic mood constraint} for echo loop prevention in same-domain agent meshes (MMP~\S8.2, \S15.8), general beyond music; (4)~the \textbf{first production deployment} of MMP/SVAF on consumer mobile hardware, with strict v0.2.2 conformance enforced by the shipping SDK; and (5)~a \textbf{deployed, verifiable system} shipping in the App Store on real listener devices, with publicly available SDKs and protocol specification. Music is the case study; the substrate is the contribution.

The quantitative deployment evaluation --- controlled comparisons against a reactive ablation and standard sequential-recommender baselines, generalisation across listeners, PAF divergence between users on the same tracks, and the first measurements of mesh-runtime CfC coherence on co-listening sessions --- is the subject of an extended companion paper currently in preparation. We have specified the evaluation protocol in full in \S\ref{sec:evaluation} so that the present architectural results can be interpreted in context and so that the protocol itself can be reviewed and refined in advance.

We believe the most useful immediate contribution is the demonstration that a continuous-time, two-cognition-layer architecture with structured CMB exchange, deployed on-device and at production scale of one shipping application, is now a practical engineering choice rather than a research curiosity. The components are small. The latency budget fits inside a mainstream mobile application's inner loop. The mesh substrate exists, is published, and is enforced by the shipping code. The remaining work --- the controlled evaluation, the history-aware encoder, the closed-loop mesh-influenced curator, additional non-music agent applications --- is incremental, not foundational. The architecture is the load-bearing result, and the architecture is general beyond music.

\subsection*{Use of AI Tools}
Claude (Anthropic) was used for drafting assistance and language editing. The author is responsible for all scientific content, experimental design, claims, and any errors.


\appendix
\section{Full PAF Behavioral Log}
\label{sec:appendix-paf}

Table~\ref{tab:paf-full} lists all 46 behavioral observations from the live deployment session described in \S\ref{sec:paf-evidence}.

\begin{table*}[h]
\centering
\caption{Full PAF behavioral log (46 observations, 12 April 2026, afternoon--evening).}
\label{tab:paf-full}
\small
\begin{tabular}{rllllrrrr}
\toprule
\# & Genre & Time & Signal & Pos & MEI\textsubscript{a} & UEA\textsubscript{a} & $\Delta$ \\
\midrule
1 & pop & aft & skipMid & 0 & $+0.40$ & $+0.10$ & $+0.30$ \\
2 & pop & aft & skipEarly & 2 & $-0.28$ & $+0.10$ & $-0.38$ \\
3 & jazz & aft & completed & 0 & $-0.48$ & $+0.00$ & $-0.48$ \\
4 & jazz & aft & skipMid & 0 & $-0.48$ & $-0.52$ & $+0.04$ \\
5 & house & aft & completed & 0 & $+0.04$ & $-0.01$ & $+0.05$ \\
6 & dance & aft & skipMid & 3 & $-0.16$ & $-0.01$ & $-0.15$ \\
7 & pop & aft & skipMid & 0 & $-0.27$ & $-0.01$ & $-0.26$ \\
8 & pop & aft & skipEarly & 1 & $+0.31$ & $-0.01$ & $+0.32$ \\
9 & hard\_rock & aft & skipEarly & 0 & $-0.04$ & $+0.00$ & $-0.04$ \\
10 & jazz & aft & completed & 0 & $-0.30$ & $+0.00$ & $-0.30$ \\
11 & classic\_rock & aft & completed & 0 & $+0.30$ & $+0.00$ & $+0.30$ \\
12 & hard\_rock & aft & completed & 0 & $+0.76$ & $+0.76$ & $+0.00$ \\
13 & jazz & aft & completed & 0 & $-0.30$ & $+0.00$ & $-0.30$ \\
14 & classical & aft & completed & 0 & $-0.30$ & $+0.00$ & $-0.30$ \\
15 & classical & aft & completed & 1 & $-0.16$ & $+0.00$ & $-0.16$ \\
16 & electronic & aft & skipMid & 3 & $+0.22$ & $+0.00$ & $+0.22$ \\
17 & pop & aft & skipEarly & 0 & $+0.04$ & $+0.21$ & $-0.17$ \\
18 & pop & aft & skipEarly & 1 & $+0.20$ & $+0.21$ & $-0.01$ \\
19 & pop & aft & skipEarly & 2 & $+0.20$ & $+0.21$ & $-0.01$ \\
20 & pop & aft & skipEarly & 3 & $-0.28$ & $+0.21$ & $-0.49$ \\
21 & pop & aft & skipEarly & 4 & $+0.10$ & $+0.21$ & $-0.11$ \\
22 & pop & aft & skipMid & 0 & $+0.04$ & $+0.00$ & $+0.04$ \\
23 & pop & aft & skipEarly & 1 & $+0.20$ & $+0.00$ & $+0.20$ \\
24 & pop & aft & skipEarly & 3 & $-0.28$ & $+0.00$ & $-0.28$ \\
25 & pop & aft & skipEarly & 4 & $+0.36$ & $+0.00$ & $+0.36$ \\
26 & pop & aft & skipEarly & 5 & $-0.08$ & $+0.00$ & $-0.08$ \\
27 & pop & aft & skipEarly & 6 & $-0.08$ & $+0.00$ & $-0.08$ \\
28 & pop & aft & skipMid & 7 & $+0.14$ & $+0.00$ & $+0.14$ \\
29 & pop & aft & skipEarly & 8 & $+0.22$ & $+0.00$ & $+0.22$ \\
30 & pop & aft & skipEarly & 9 & $-0.08$ & $+0.00$ & $-0.08$ \\
31 & pop & aft & skipEarly & 10 & $-0.14$ & $+0.00$ & $-0.14$ \\
32 & pop & aft & skipEarly & 11 & $+0.30$ & $+0.00$ & $+0.30$ \\
33 & pop & aft & skipEarly & 12 & $-0.08$ & $+0.00$ & $-0.08$ \\
34 & pop & aft & skipEarly & 13 & $+0.32$ & $+0.00$ & $+0.32$ \\
35 & electronic & eve & completed & 0 & $-0.40$ & $+0.10$ & $-0.50$ \\
36 & electronic & eve & completed & 0 & $-0.16$ & $+0.10$ & $-0.26$ \\
37 & electronic & eve & completed & 1 & $+0.44$ & $+0.10$ & $+0.34$ \\
38 & electronic & eve & completed & 2 & $+0.40$ & $+0.10$ & $+0.30$ \\
39 & electronic & eve & volumeDown & 4 & $-0.10$ & $+0.10$ & $-0.20$ \\
40 & electronic & eve & volumeUp & 4 & $-0.10$ & $+0.10$ & $-0.20$ \\
41 & Cantopop & eve & completed & 0 & $-0.30$ & $+0.00$ & $-0.30$ \\
42 & Mandopop & eve & volumeUp & 5 & $-0.30$ & $+0.00$ & $-0.30$ \\
43 & Mandopop & eve & volumeUp & 5 & $-0.30$ & $+0.00$ & $-0.30$ \\
44 & Mandopop & eve & volumeDown & 5 & $-0.30$ & $+0.00$ & $-0.30$ \\
45 & dance & eve & completed & 0 & $+0.04$ & $+0.18$ & $-0.14$ \\
46 & edm & eve & completed & 0 & $+0.70$ & $+0.42$ & $+0.28$ \\
\bottomrule
\end{tabular}
\end{table*}

\end{document}